\documentclass[12pt]{article}
\usepackage{epsfig}
\usepackage{axodraw}

\def\theequation{\thesection.\arabic{equation}}

\def\simlt{\stackrel{<}{{}_\sim}}
\def\simgt{\stackrel{>}{{}_\sim}}

\newcommand{\bea}{\begin{eqnarray}}
\newcommand{\eea}{\end{eqnarray}}
\newcommand{\bd}{\begin{displaymath}}
\newcommand{\ed}{\end{displaymath}}
\newcommand{\be}{\begin{equation}}
\newcommand{\ee}{\end{equation}}

\def\theequation{\thesection.\arabic{equation}}

\renewcommand{\baselinestretch}{1.2}

\begin{document}

\thispagestyle{empty}

{\normalsize\sf
\rightline {hep-ph/0110249}
\rightline{IFT-01/27}
}

\vskip 5mm

\begin{center}
  
{\LARGE\bf Quantum corrections to neutrino masses and mixing 
angles}\footnote{To appear in the review section of Modern Physics A.}

\vskip 10mm

{\large\bf Piotr H. Chankowski and Stefan Pokorski}\\[5mm]

Institute of Theoretical Physics, Warsaw University\\
Ho\.za 69, 00-681 Warsaw, Poland

\end{center}

\vskip 5mm

\renewcommand{\baselinestretch}{1.1} 
\begin{center}
{\bf Contents}
\end{center}
{\small
\begin{enumerate}
\item[1] Introduction
\item[2] Fermion masses and mixing
\item[3] Neutrino masses in the effective theory
\item[4] Quantum corrections from the renormalization group evolution
  \begin{enumerate}
  \item[4.1] RG equations for the CKM matrix
  \item[4.2] RG equations for neutrino masses and mixing angles
  \item[4.3] Evolution of the neutrino masses
  \item[4.4] Mixing of two neutrinos
  \item[4.5] Mixing of three neutrinos and fixed points
  \end{enumerate}
\item[5] Low energy threshold corrections
  \begin{enumerate}
  \item[5.1] Threshold corrections in the SM
  \item[5.2] Threshold corrections in the MSSM
    \begin{enumerate}
    \item[5.2.1] Three-fold degeneracy and flavour diagonal corrections
    \item[5.2.2] Three-fold degeneracy and flavour non-diagonal corrections
    \item[5.2.3] Two-fold degeneracy and threshold corrections
    \end{enumerate}
  \end{enumerate}
\item[6] Conclusions
\end{enumerate}
\noindent Appendix A\\
\noindent Appendix B\\
\noindent References
}
\renewcommand{\baselinestretch}{1.2}

\newpage \setcounter{page}{1} \setcounter{footnote}{0}
\section{Introduction}
\setcounter{equation}{0} 

Quark and lepton masses and mixing angles are free parameters of the Standard 
Model (SM). They are known with various degree of accuracy, from the precision 
direct measurements of charged lepton masses, a combination of experimental 
data and theoretical arguments for quark masses and relatively precise 
measurements of quark mixing angles and the phase parameterizing the violation 
of CP invariance in the SM. A new element in this picture are neutrino masses 
and mixing angles. There is at present strong experimental evidence for 
neutrino oscillations whose most obvious and most natural explanation is that 
neutrinos have non-zero masses and the neutrino mass eigenstates are different 
{}from the weak interaction eigenstates. Although far from having final 
interpretation, the present experimental data give interesting preliminary 
information about the neutrino mass sector and the forthcoming experiments are
expected to resolve the remaining ambiguities. 
 
On the theoretical side, the origin of the interactions giving rise to 
fermion masses is a problem that cannot be addressed in the framework of 
the SM. The physical scale (we shall call it $M$) of the, still unknown, 
theory of fermion masses is certainly above the electroweak scale and quite 
likely it may even be close to the GUT or Planck scales. Such a high scale
is suggested, for instance, by the see-saw interpretation of the magnitude
of the neutrino masses indicated by experiment. Ultimately, the
theory will predict at the scale $M$ the running fermion mass parameters 
(and perhaps also other parameters) of the effective low energy theory, 
describing physics at energies $< M$.
\footnote{The running mass parameters of the effective low energy theory 
can in principle be calculated in the underlying theory at any renormalization 
scale $Q$ by proper inclusion of the high energy threshold corrections.
Choosing $Q\approx M$ one minimizes those corrections.}

To relate the mass parameters of the effective low energy theory to the 
experimentally measured quantities one has to include quantum 
corrections that already do not depend on the specific theory of fermion 
masses but only on the low energy effective theory. Clearly, close to the 
electroweak scale this theory is the Standard Model (SM). One possibility is 
that the SM remains the correct effective theory up to the scale $M$. This 
is conceivable particularly if the scale $M$ is relatively low. Another 
possibility is that the SM needs to be embedded into a bigger effective 
theory already much below $M$. The latter case may, for instance, happen if 
the low energy supersymmetry is realized in Nature. In this review we discuss 
quantum corrections to neutrino masses both in the SM and in its 
supersymmetric extension, the Minimal Supersymmetric Standard Model (MSSM).

Quite generally one can distinguish two classes of quantum corrections that 
enter on a somewhat different footing. The first one is given by the RGEs 
describing the evolution of the fermion mass parameters in the SM or MSSM 
{}from the scale $M$ down to some scale close to the electroweak scale. Another
source of quantum corrections are the so-called low energy threshold effects.
Strictly speaking these are the corrections necessary to express measurable
quantities like neutrino masses and mixing angles in terms of the running
(renormalized) parameters of the effective theory Lagrangian. Formally, their 
inclusion renders the prediction for observables independent of the choice of 
the scale to which the RGEs are integrated. In the SM the threshold 
corrections are unambiguous 
and, if that scale is taken to be close to $M_Z$, they are small and can be 
neglected. Such corrections may, however, be very important in the MSSM
since there they depend also on the sparticle masses and couplings.  

In Section 2 we review the neutrino masses and mixing and stress the 
differences with the quark sector. Contrary to the small mixing in the 
latter one, at least one neutrino mixing angle and, quite likely two, are 
close to maximal \cite{NI}. A maximal or bimaximal mixing would be quite 
natural for (approximately) degenerate masses, much larger than their 
differences \cite{ALFE}. Such 
mass patterns are very different from the hierarchical masses of the 
charged fermions but are consistent with experiment. Indeed, only mass 
squared differences can be inferred from the data and not the neutrino 
masses themselves. 

The observed (mixing) and potential (mass pattern) differences between the
neutrino and quark sectors provide strong motivation for studying quantum
corrections to neutrino masses and mixing. It is well known that the 
corrections are small for the quarks, just because of their hierarchical 
masses and mixing. As we shall review in this paper, that remains true for
neutrinos if their masses are hierarchical too. However, quantum corrections
may give strong, qualitatively new, effects if the neutrino masses are not
hierarchical. 

The main parts of this review are Sections 4 and 5. In Section 4 we discuss 
quantum corrections described by the renormalization group evolution and in 
Section 5 the potential effects of the low energy threshold corrections are 
reviewed.

A brief overview of quantum corrections in the neutrino sector is given in
Section 6.

\section{Fermion masses and mixing}
\setcounter{equation}{0} 

For quarks and charged leptons the Particle Data Group \cite{PDG} gives the 
following values of the masses\footnote{For $u$, $d$ and $s$ quarks the quoted 
values are the running masses in the $\overline{\rm MS}$ scheme at $Q=2$ 
GeV. The values given for the $c$ and $b$ quarks are the $\overline{\rm MS}$ 
running masses at $Q=m_c$ and  $Q=m_b$, respectively. Finally $m_t$ given 
here is the pole mass.}:
\begin{eqnarray}
m_u=1.5-5.0 {\rm ~MeV},\phantom{aaa}m_c=1.15-1.35 {\rm ~GeV},
\phantom{aaa}m_t=174.3\pm5.1 {\rm ~GeV},\nonumber\\
m_d=3.0-9.0 {\rm ~MeV},\phantom{aaaa}m_s=60-170 {\rm ~MeV},
\phantom{aaaa}m_b=4.0-4.4 {\rm ~GeV},\phantom{a}\\
m_e=511 {\rm ~keV},\phantom{aaaaa}m_\mu=105.7 {\rm ~MeV},
\phantom{aaaaa}m_\tau=1.777 {\rm ~GeV}\phantom{aaaa}\nonumber
\end{eqnarray}
and of the quark mixing (absolute values of the elements of the 
Cabibbo-Kobayashi-Maskawa (CKM) matrix):
\begin{eqnarray}
\left(\matrix{0.9742-0.9757&0.219-0.226&0.002-0.005\cr
             0.219-0.225&0.9743-0.9749&0.037-0.043\cr
             0.004-0.014&0.035-0.043&0.9990-0.9993}\right)
\end{eqnarray}
The pattern of quark masses and mixing is clear: hierarchical masses and 
small (and also) hierarchical mixing: the larger the mass difference, the 
smaller the mixing angle. In the framework of the SM, fermion masses and 
mixing angles are used to fix the Yukawa matrices which are the actual free 
parameters of the theory. They cannot, however, be uniquely reconstructed from
the experimental data: there is always the ambiguity of rotating the 
electroweak basis and all we can have at present is merely a phenomenological 
parametrization. 

Neutrino masses and mixing are inferred from a number of oscillation
experiments. The interpretation of neutrino oscillations in terms of massive 
neutrinos is the most natural one. To discuss this interpretation, let 
us first recall the oscillation pattern for two hypothetical neutrinos with 
masses $m_1$ and $m_2$ whose quantum fields are linear combinations of two 
quantum fields $\nu_A$ and $\nu_B$ that are weak eigenstates
\begin{eqnarray}
&&\nu_1 = \cos\vartheta\nu_A + \sin\vartheta\nu_B\nonumber\\
&&\nu_2 = -\sin\vartheta\nu_A + \cos\vartheta\nu_B
\label{eqn:mix}
\end{eqnarray}
The probabilities that in a given experiment a neutrino produced in the 
interaction of the charged lepton of flavour $A$  with the $W^\pm$ boson is
detected at a distance $L$ as the neutrino creating the charged lepton of 
flavour $B$, $B\neq A$, or $A$ are\footnote{The formulae like 
(\ref{eqn:2x2mix}) are usually derived in the framework of quantum mechanics. 
The proper picture of neutrino oscillations is however the field theoretical 
one \cite{CA}. Due to the field mixing (\ref{eqn:mix}) there is in general a 
nonzero amplitude for emission (absorption) of any of the neutrino mass 
eigenstates $\nu_a$ in the interaction of charged lepton of flavour $A$ with 
the $W$-boson. The change of the neutrino flavour (inferred from the flavours 
of the charged leptons) is in this picture due to the coherent sum of Feynman 
diagrams describing exchanges of all virtual mass eigenstates of neutrinos 
between the emission and detection vertices. Taking properly into account the 
effects of wave packets describing initial and final states it can be shown 
that the formula (\ref{eqn:2x2mix}) is in most cases a sufficient 
approximation to the full result, which automatically accounts for decoherence 
effects and depends on the overlap of the wave packets describing the
initial and final states \cite{CA}.}
\begin{eqnarray}
&&P^{2\times2}(\nu_A\rightarrow\nu_B) = \sin^22\vartheta
\sin^2\left({\Delta m^2L_{\rm exp}\over4E_{\rm exp}}\right)\nonumber\\
&&P^{2\times2}(\nu_A\rightarrow\nu_A) = 1-P^{2\times2}(\nu_A\rightarrow\nu_B)
\label{eqn:2x2mix}
\end{eqnarray}
We see that the oscillation probability depends on 
$\Delta m^2\equiv m^2_{\nu_2}-m^2_{\nu_1}$, the distance $L_{\rm exp}$, the 
mixing angle $\vartheta$ and the neutrino energy $E_{\rm exp}$. It is clear 
that for observing the oscillation pattern the factor $\Delta m^2L/4E$ should 
be of order ${\cal O}(1)$. \footnote{Recall 
that ${\Delta m^2L\over4E}=1.27\times{(\Delta m^2/1 ~{\rm eV}^2)
(L/1 ~{\rm km})\over(E/1 ~{\rm GeV})}$.} 
Hence, in general, longer oscillation distances are necessary to probe smaller 
mass squared differences. On the other hand, for $\Delta m^2L/4E\gg1$ the 
transition probability is 
$P(\nu_A\rightarrow\nu_B)\approx{1\over2}\sin^22\vartheta$, i.e. it is 
insensitive to $\Delta m^2$, because in realistic applications the 
expression (\ref{eqn:2x2mix}) has to be averaged over some non-zero interval 
of the initial neutrino energies $\Delta E$ \cite{BIGIGR}.

We can now summarize the results of various experiments. They are 
usually interpreted in terms of the effective $2\times2$ parametrization 
(\ref{eqn:2x2mix}). Historically, the first information came from the 
experiments measuring the flux of $\nu_e$ neutrinos produced in the Sun (for 
references see e.g. the review \cite{BIGIGR}). A convincing evidence for the
solar neutrino oscillation was provided by the Kamiokande and Superkamiokande 
experiments which established a strong ($\approx50$\%) suppression of the 
flux of neutrinos from the nuclear reaction 
$^8B\rightarrow\phantom{a}^8Be^\ast+e^++\nu_e$. The most plausible 
explanation of those results is the transmutation of $\nu_e$'s produced in 
the core of the Sun into another type of neutrinos such as $\nu_\mu$, 
$\nu_\tau$ and/or the so-called sterile neutrino $\nu_{\rm sterile}$ which 
does not interact with the $W^\pm$ or $Z^0$ bosons. Such a transmutation can 
occur either during their flight from 
the Sun to the Earth (the so-called vacuum oscillations (VO)) or through the 
resonant transition in the matter of the outer layers of the Sun (the 
so-called MSW effect \cite{MISMWO}). As follows from the formulae 
(\ref{eqn:2x2mix}) with $A\equiv e$ and $B\equiv\mu$, $\tau$ or $s$, in the 
case of VO with $L\approx1.5\times10^8$~km (the Sun-Earth mean distance) and 
for mean ~$^8B$ neutrino energy of order $E\sim10$~MeV, the solar neutrino 
experiments are sensitive to 
$\Delta m^2_{\rm sol}\sim{\cal O}(10^{-(11-10)})$~eV$^2$. The deficit is then 
explained with $\sin^22\vartheta_{\rm sol}>0.7$. In the case of the MSW 
resonant conversion the formulae (\ref{eqn:2x2mix}) for the transition and 
survival probabilities are replaced by more complicated expressions 
\cite{BIGIGR} which depend also on the electron and neutron number densities 
in the Sun. The observed $\nu_e$ neutrino deficit can be then explained 
either for $\Delta m^2_{\rm sol}\sim{\cal O}(10^{-5})$~eV$^2$ and 
$\sin^22\vartheta_{\rm sol}\sim{\cal O}(10^{-(3-2)})$ (the so-called SAMWS
solution) or for $\Delta m^2_{\rm sol}\sim{\cal O}(10^{-4})$~eV$^2$ and 
$\sin^22\vartheta_{\rm sol}>0.5$ (the so-called LAMWS solution).

The deficit of neutrinos was also revealed by the measurements of the flux 
of $\nu_\mu$ and $\bar\nu_\mu$ neutrinos produced together with $\nu_e$'s in 
the Earth's atmosphere by the cosmic rays. The results of the Superkamiokande 
experiment \cite{SUPERK} are most easily explained by $\nu_\mu$ oscillation 
into another type of neutrinos. The oscillatory explanation is further 
supported by the zenith angle dependence of the $\nu_\mu$ and $\bar\nu_\mu$ 
flux deficit. For typical $\nu_\mu$ energies of order $\sim$GeV and 
20~km $\simlt L\simlt1.3\times10^4$~km, the observed $\nu_\mu$ deficit is 
explained for $\Delta M^2_{\rm atm}\approx3.2\times10^{-3}$~eV$^2$ and 
$\sin^22\vartheta_{\rm atm}>0.82$. Also, the Superkamiokande data
seem to favour $\nu_\mu\rightarrow\nu_\tau$ oscillations over 
$\nu_\mu\rightarrow\nu_{\rm sterile}$ \cite{SUPERK_NOs}.

Neutrino oscillations are also intensively searched for in various reactor or 
accelerator based experiments (for review see \cite{BIGIGR,BEGRVO}). Except 
for the LSND experiment reporting \cite{LSND} a positive signal for the 
$\bar\nu_\mu\rightarrow\bar\nu_e$ oscillations, which would require
$0.1~{\rm eV}^2\simlt\Delta M^2\simlt1~{\rm eV}^2$ for the oscillatory 
explanation, the results of all other experiments are negative. Particularly 
strong constraint comes from the CHOOZ reactor experiment. It excludes 
disappearance 
of $\bar\nu_e$ with mean energies $\sim3$~MeV at the distance $L\approx1$~km.
In the $2\times2$ oscillation framework this translates into the limit
\begin{eqnarray}
\sin^22\vartheta_{\rm react}<0.1 ~~~{\rm for} ~~~
\Delta M^2\simgt10^{-3}~{\rm eV}^2
\label{eqn:CHOOZ}
\end{eqnarray}

Except for the unconfirmed LSND result, all experiments are consistent 
with oscillations between the three known neutrino flavours. Moreover,
recent SNO measurement \cite{SNO} of the total solar neutrino flux (i.e. the 
flux of $\nu_e$, $\nu_\mu$ and $\nu_\tau$) combined with the Superkamiokande 
data \cite{SUPERK_sol} strongly supports this assumption. Therefore, in the 
rest of the review we assume that the neutrino sector of the low energy 
effective theory consists of 3 active neutrinos only. 

Within the $3\times3$ framework the
solar, atmospheric and reactor data become interrelated and one can draw
more definite conclusions. To make the discussion of the neutrino experiments
more complete we recall the formula for the transition probabilities 
$P(\nu_A\rightarrow\nu_B)$ for the case of 3 neutrinos:
\begin{eqnarray}
&&P(\nu_A\rightarrow\nu_B,L,E)=
\left|\sum_aU_{Ba}e^{-i{m^2_{\nu_a}L\over2E}}U^\ast_{Aa}\right|^2\nonumber\\
&&\phantom{aaaaaa}=\sum_a\left|U_{Ba}U^\ast_{Aa}\right|^2
+{\rm Re}\sum_a\sum_{b\neq a}U_{Ba}U^\ast_{Bb} U^\ast_{Aa}U_{Ab}
e^{i{\Delta m^2_{ab}L/4E}}
\phantom{aaa}\label{eqn:pdb}
\end{eqnarray}
where $\Delta m^2_{ab}\equiv m^2_{\nu_a}-m^2_{\nu_b}$. The complex matrix 
elements are defined in the basis in which charged lepton mass matrix is 
diagonal, by the decomposition of the $\nu_A$ neutrino fields belonging to 
$SU_L(2)$ doublets into the mass eigenstates field:
\begin{eqnarray}
\nu_A = \sum_aU_{Aa}\nu_a.
\end{eqnarray}
For a real matrix $U$ the expression (\ref{eqn:pdb}) can be written as
\begin{eqnarray}
P(\nu_A\rightarrow\nu_B,L,E)=
\delta_{AB}-4\sum^3_{a>b=1}
U_{Aa}U_{Ba}U_{Ab}U_{Bb}\sin^2\left({\Delta m^2_{ab}L\over4E}\right).
\end{eqnarray}

To simplify the notation we will denote: 
$\Delta m^2\equiv\Delta m^2_{\rm sol}\equiv |m^2_2-m^2_1|$, 
$\Delta M^2\equiv\Delta M^2_{\rm atm}\equiv |m^2_3-m^2_2|$. 
For $\Delta m^2\ll\Delta M^2$ the 
measurements of the experiments listed above can now be summarized as follows:
\begin{eqnarray}
P_{\rm sol}(\nu_e\rightarrow\nu_e)&=&
1-2\left(1-U^2_{13}\right)U^2_{13}-4U^2_{11}U^2_{12}
\sin^2\left({\Delta m^2L_{\rm sol}\over4E_{\rm sol}}\right)\nonumber\\
P_{\rm atm}(\nu_\mu\rightarrow\nu_\mu)&=&
1-4\left(1-U^2_{23}\right)U^2_{23} 
\sin^2\left({\Delta M^2L_{\rm atm}\over4E_{\rm atm}}\right)\\
P_{\rm react}(\nu_e\rightarrow\nu_e) &=& 
1-4\left(1-U^2_{13}\right)U^2_{13} 
\sin^2\left({\Delta M^2L_{\rm react}\over4E_{\rm react}}\right)\nonumber
\end{eqnarray}
A convenient parametrization for the mixing matrix $U$ is \cite{MANASA}
(see also the next Section):
\begin{eqnarray}
U_{\rm MNS}=\left(\matrix{c_{12}c_{13} & s_{12}c_{13} & s_{13}e^{-i\delta}\cr
-s_{12}c_{23}-c_{12}s_{23}s_{13}e^{i\delta} 
&c_{12}c_{23}-s_{12}s_{23}s_{13}e^{i\delta} & s_{23}c_{13}\cr
s_{12}s_{23}-c_{12}c_{23}s_{13}e^{i\delta} 
&-c_{12}s_{23}-s_{12}c_{23}s_{13}e^{i\delta} & c_{23}c_{13}}\right)\times\Phi
\label{eqn:MNSmat}
\end{eqnarray}
where $c_{ij}\equiv\cos\theta_{ij}$ ($s_{ij}\equiv\sin\theta_{ij}$) and $\Phi$
is a diagonal matrix $\Phi=diag(e^{i\alpha_1},e^{i\alpha_2},1)$ (or a similar
diagonal matrix with two phases and $\Phi_{11}=1$ or $\Phi_{22}=1$). Our
convention is such that for vanishing mixing angles $\theta_{ij}$ we have
$\nu_1=\nu_e$, $\nu_2=\nu_\mu$ and $\nu_3=\nu_\tau$, i.e. no ordering of
the neutrino masses $m_{\nu_a}$ is assumed.
CP is conserved if $\delta=0$ mod $\pi$ and $\alpha_{1,2}=0$ mod $\pi/2$. 
The relation to the effective $2\times2$ parametrization then reads
\begin{eqnarray}
\sin^22\vartheta_{\rm react}&=&\sin^22\theta_{13}\nonumber\\
\sin^22\vartheta_{\rm atm}&=&\sin^22\theta_{23}\left(\cos^2\theta_{13}
+{\sin^22\theta_{13}\over4\cos^2_{23}}\right)\nonumber\\
\sin^22\vartheta_{\rm sol}&\approx&\sin^22\theta_{12}\left(1-
2\sin^2\theta_{13}\right)
\end{eqnarray}
In the limit of small $\theta_{13}$ angle imposed by the CHOOZ result
(\ref{eqn:CHOOZ}) (non-negligible entries $U_{23}$ and $U_{33}$ and unitarity 
of $U$ exclude the possibility of $|\theta_{13}|\approx\pi/2$) we get
\begin{eqnarray}
\sin^22\vartheta_{\rm atm}&\approx&\sin^22\theta_{23}\nonumber\\
\sin^22\vartheta_{\rm sol}&\approx&\sin^22\theta_{12}
\end{eqnarray}

Finally we mention that the neutrino Majorana mass term of 
the form ${\cal L}_{\rm mass}=-{1\over2}m_\nu^{AB}\nu_A\nu_B+{\rm H.c}$ 
violates the lepton number conservation and leads to the neutrino-less
double beta decay. Non-observation of such decays implies \cite{BETA0NIU}
\begin{eqnarray}
|m_\nu^{ee}|<0.35 (0.27) {\rm ~eV ~~~at ~90\% ~(68\%)~C.L.},
\end{eqnarray}
i.e.
\begin{eqnarray}
|m_{\nu_1}c^2_{12}c^2_{13}e^{2i\alpha_1} + 
m_{\nu_2}s^2_{12}c^2_{13}e^{2i\alpha_2} + m_{\nu_3}s^2_{13}e^{2i\delta}|
<0.35 (0.27) {\rm ~eV ~~at ~90\% ~(68\%)~C.L.}
\label{eqn:2bb0nubound}
\end{eqnarray}

Concluding this section,  we stress that the pattern of neutrino mixing is 
distinctly different from the one observed in the quark sector. At least one 
mixing angle ($\theta_{23}$) is large and the recent data favour solutions
with two large mixing angles \cite{NI}. The data indicate 
$\Delta m^2\ll\Delta M^2$ but, since the masses are not measured,  
several options for the values of the masses themselves are still possible:
\begin{itemize}
\item[$i)$] $\Delta M^2\approx m^2_{\nu_3}
            \gg m^2_{\nu_{2(1)}}\gg m^2_{\nu_{1(2)}}$ 
            (hierarchical)
\item[$ii)$] $\Delta M^2\approx m^2_{\nu_1}, m^2_{\nu_2}\gg m^2_{\nu_3}$ or
             $\Delta M^2\approx m^2_{\nu_3}\gg m^2_{\nu_1}, m^2_{\nu_2}$, with
             $m^2_{\nu_1}, m^2_{\nu_2}\gg\Delta m^2$ (partly degenerate)
\item[$iii)$] $m^2_{\nu_3}\approx m^2_{\nu_2}\approx m^2_{\nu_1}$ and all of 
             order or larger  than $\Delta M^2$ (degenerate). 
\end{itemize}
Special cases of the partly degenerate and degenerate patterns are equal 
masses, $|m_{\nu_1}|=|m_{\nu_2}|$ or $|m_{\nu_1}|=|m_{\nu_2}|=|m_{\nu_3}|$,
respectively. In our language we refer to them as two-fold and three-fold
degeneracies, to distinguish them from, more general, partly degenerate and
degenerate pattern defined by $ii)$ and $iii)$.
The last two possibilities are very different from the pattern known from 
the quark sector but almost maximal (bimaximal) mixing suggested by the
experimental data makes them an interesting alternative to the hierarchical
pattern. It is important to remember that in that respect neutrinos may be
qualitatively different from quarks because, unlike the charged fermions, they 
can be Majorana particles. As we shall see, the magnitude and the importance 
of quantum corrections depends strongly on the assumed pattern of the masses.

\section{Neutrino masses in the effective theory}
\setcounter{equation}{0} 

We begin by recalling how the quark masses are incorporated in the SM (or the
MSSM) Lagrangian. We shall use the Weyl spinor notation (see e.g. 
\cite{POKBOOK}), which is particularly convenient for Majorana particles.
The relation to the standard Dirac notation is explained in the Appendix A.
The original Lorentz and $SU(3)\times SU(2)\times U(1)$ invariant
Lagrangian contains the Yukawa interactions of the form
\begin{eqnarray}
{\cal L}_{\rm Yuk}=
-\epsilon_{ij}H_i u^{cA}\mbox{\boldmath$Y$}_u^{AB}q^B_j
-H_j^\ast d^{cA}\mbox{\boldmath$Y$}_d^{AB}q^B_j
-H_j^\ast e^{cA}\mbox{\boldmath$Y$}_e^{AB}l^B_j+{\rm H.c.}
\label{eqn:smlyuk}
\end{eqnarray}
where $\mbox{\boldmath$Y$}_u$, $\mbox{\boldmath$Y$}_d$, 
$\mbox{\boldmath$Y$}_e$ are a priori arbitrary complex Yukawa matrices, 
$\epsilon_{12}=-\epsilon_{21}=-1$, the upper case letters enumerate the 
three generations of the matter fermions
\begin{eqnarray}
u^c, ~~d^c, ~~e^c, ~~q\equiv\left(\matrix{u\cr d}\right) ~~{\rm and} ~~
l\equiv\left(\matrix{\nu\cr e}\right) 
\label{eqn:smfermions}
\end{eqnarray}
described by the left-handed Weyl spinors and $H$ is the SM Higgs doublet.
In the MSSM there are two Higgs doublets, $H^{(u)}$ and 
$H^{(d)}$, and one has to replace: $H_i\rightarrow H^{(u)}_i$ and
$H_j^\ast\rightarrow\epsilon_{ij}H_i^{(d)}$. Replacing the Higgs doublet $H$ 
(doublets $H^{(u)}$ and $H^{(d)}$ in the MSSM) by its vacuum expectation value
leads to the following mass terms in the SM (MSSM) Lagrangian:
\begin{eqnarray}
{\cal L}_{\rm mass}=
-u^{cA}(v\mbox{\boldmath$Y$}_u)^{AB}u^B
-d^{cA}(v\mbox{\boldmath$Y$}_d)^{AB}d^B 
-e^{cA}(v\mbox{\boldmath$Y$}_e)^{AB}e^B+{\rm H.c.}
\label{eqn:smlmass}
\end{eqnarray}
(In the MSSM $v\mbox{\boldmath$Y$}_u\rightarrow v_u\mbox{\boldmath$Y$}_u$,
$v\mbox{\boldmath$Y$}_{d,e}\rightarrow -v_d\mbox{\boldmath$Y$}_{d,e}$.)
The fermion mass matrices are diagonalized by the unitary chiral rotations
\begin{eqnarray}
&&u\rightarrow \mbox{\boldmath$U$}_L u, ~~~
u^c\rightarrow u^c \mbox{\boldmath$U$}_R^\dagger,\nonumber\\
&&d\rightarrow \mbox{\boldmath$D$}_L d, ~~~
d^c\rightarrow d^c \mbox{\boldmath$D$}_R^\dagger,
\phantom{aaa}\label{eqn:rotations}\\
&&e\rightarrow \mbox{\boldmath$E$}_L e, ~~~
e^c\rightarrow e^c \mbox{\boldmath$E$}_R^\dagger,
\nonumber
\end{eqnarray}
which give:
\begin{eqnarray}
\mbox{\boldmath$U$}_R^\dagger\mbox{\boldmath$Y$}_u\mbox{\boldmath$U$}_L 
={\rm diag}(y_u, ~y_c, ~y_t)\nonumber\\
\mbox{\boldmath$D$}_R^\dagger\mbox{\boldmath$Y$}_d\mbox{\boldmath$D$}_L 
={\rm diag}(y_d, ~y_s, ~y_b)\label{eqn:diagyuk}\\
\mbox{\boldmath$E$}_R^\dagger\mbox{\boldmath$Y$}_e\mbox{\boldmath$E$}_L 
={\rm diag}(y_e, ~y_\mu, ~y_\tau).\nonumber
\end{eqnarray}
The only remnant of the nontrivial Yukawa matrices is then the CKM matrix
\begin{eqnarray}
V_{\rm CKM}\equiv V\equiv\mbox{\boldmath$U$}_L^\dagger\mbox{\boldmath$D$}_L
\label{eqn:ckmmatrix}
\end{eqnarray}
appearing in the interactions of the charged quark currents with (massive) 
charged vector bosons $W^\pm_\mu$. It can be shown (see e.g. \cite{POKBOOK})
that the CKM matrix can be parameterized by 3 angles and one phase. The other 
$6-1=5$ phases can be absorbed in the redefinition of the quark fields.

In the SM defined as a renormalizable theory, with the neutrino and Higgs 
fields transforming as components of doublets of the $SU_L(2)$ gauge 
symmetry, neutrinos remain massless. Moreover, as a consequence of the gauge
symmetry, renormalizability and the field content, the Lagrangian has two
accidental global $U(1)$  symmetries that ensure baryon and lepton number
conservation. In the absence of any mass terms for the neutrino fields 
$\nu^A$, one can always perform the rotation 
$\nu\rightarrow\mbox{\boldmath$V$}_L\nu$ with 
$\mbox{\boldmath$V$}_L=\mbox{\boldmath$E$}_L$ so that the counterpart of
the CKM matrix in the leptonic sector is trivial.

There are two easy possibilities for extending the particle content of the
SM so that neutrino mass is generated by renormalizable interactions. One
is to couple two lepton $SU_L(2)$ doublets to a $SU_L(2)$ Higgs triplet 
(singlets would violate the electric charge conservation). This possibility 
is not particularly attractive for several reasons. The smallness of the 
neutrino masses would require either $v_{\rm triplet}\ll v_{\rm doublet}$ 
or the triplet couplings orders of magnitude smaller than the other Yukawa 
couplings. Moreover, the introduction of the triplet would make the parameter 
$\rho$ a free parameter of the theory (hence not calculable) with its own 
counterterm allowing to adjust its value at will. The other possibility is to 
introduce a number of new $SU_L(2)\times U_Y(1)$ singlet left-handed leptonic 
fields $\nu^c$ to the Lagrangian and couple them to the Higgs doublet in the 
same way as the quark fields $u^c$: 
\begin{eqnarray}
\Delta{\cal L}_{\rm Yuk} = -\epsilon_{ij}H_i\nu^{cK} 
\mbox{\boldmath$Y$}_\nu^{KA} l^A_j + {\rm H.c.}
\label{eqn:Yukneu}
\end{eqnarray}
where  $K=1,\dots, K_{max}$ and in principle $K_{max}$ could be arbitrary.
The interaction (\ref{eqn:Yukneu}) preserves the known structure of the SM 
and, if $K=1,2,3$, the symmetry between quarks and leptons is restored. For
singlet fields $\nu^c$ the Majorana mass term 
\begin{eqnarray}
\Delta{\cal L}_{\rm Maj}=-{1\over2}\mbox{\boldmath$M$}^{KL}_{\rm Maj}
\nu^{cK}\nu^{cL}+ {\rm H.c.}
\label{eqn:Majmass}
\end{eqnarray}
can also be added to the Lagrangian. In general both terms should be included 
in the  theory defined by the $SU_L(2)\times U_Y(1)$ gauge symmetry and 
renormalizability. The Majorana mass term can be, however, eliminated  by 
imposing the additional global $U(1)$ symmetry ensuring conservation of the 
lepton number $L$. The fields $\nu^c$ must then have $L=-1$, opposite to $L$
of the leptonic doublets $l_i$. Their complex conjugate $\bar\nu^c$ (see 
Appendix A) can be then interpreted as the right-handed neutrinos. 
Neutrinos are then Dirac particles like the other fermions.
There are two reasons why Dirac masses are not so 
attractive. One is, again, the need for very small numerical values 
of the Yukawa couplings $\mbox{\boldmath$Y$}_\nu$. The other is
that the lepton number conservation has to be imposed as an additional 
global symmetry (remember that in the SM it is a consequence of the field
content and of the renormalizability and not an additional assumption).

The presence of the Majorana mass term, which breaks the global $U(1)$ 
symmetry, inevitably makes neutrinos Majorana particles (in fact what 
allows to interpret in the SM the two helicity states described by the 
$\nu$ field as a particle and an anti-particle is just the lepton number!) 
Taking for simplicity three\footnote{The formula (\ref{eqn:massmatGUT}) is 
easily generalized to an arbitrary number of singlet neutrinos $\nu^c$.}
singlet neutrinos $\nu^c$ and replacing $H$ ($H^{(u)}$ in the MSSM) by its 
vacuum expectation value, the general form of the neutrino mass matrix (in 
the absence of Higgs triplets) is
\begin{eqnarray} 
{\cal M}_\nu=\left(\matrix{0&m_D\cr m_D^T & M_{\rm Maj}}\right)
\label{eqn:massmatGUT}
\end{eqnarray}
where $m_D\equiv v\mbox{\boldmath$Y$}_\nu$ and $M_{\rm Maj}$ are $3\times3$ 
matrices in the basis $(\nu,\nu^c)$. If both these matrices are diagonal with
eigenvalues $m_A$ and $M_A$, respectively, the physical neutrinos 
are mixtures of $\nu_A$ and $\nu^c_A$ with the masses
\begin{eqnarray} 
m_{\nu^A\pm}={1\over2}\left(M_A\pm\sqrt{M_A^2+4m_A^2}\right).
\end{eqnarray}

In general, each of the six leptonic fields $\nu_A$ and $\nu^c_K$ is a linear 
combination of the six Majorana neutrinos. There is the possibility of
oscillations $\nu\rightarrow\nu^c$. The $\nu^c$ neutrinos are singlets of
the $SU_L(2)\times U(1)$ gauge group and are sterile 
neutrinos.\footnote{Note that such oscillations exist only for Majorana 
neutrinos and are forbidden by $L$ conservation for Dirac neutrinos i.e. 
the right-handed neutrinos $\bar\nu^c$ have correct $L$ but wrong 
chirality to oscillate into$\nu$.}
The situation simplifies for $M\gg v$ where $M$ is the overall scale of
the $M_{\rm Maj}$ entries. For simplicity, we identify it with the 
introduced earlier scale $M$, although they could actually be different 
scales. In this limit there are two sets of mass eigenvalues. They are
of the order ${\cal O}(M)$ and ${\cal O}(v^2/M)$, respectively, and three of 
the mass eigenstates are very heavy and decouple from the physics at the 
electroweak scale. This is the so-called see-saw mechanism \cite{GERASL}
\footnote{However in some cases, one of the formally 
${\cal O}(M)$ mass eigenvalues can remain small as a result of some symmetries
of the underlying theory of neutrino masses. The corresponding light state can 
play the role of a sterile neutrino \cite{MOH}.}
The oscillations occur then effectively among the three ``active'' neutrinos
$\nu$ (the $\nu^c$'s composition is dominated by heavy eigenstates). 

\begin{figure}[htbp]
\begin{center}
\begin{picture}(180,90)(0,0)
\ArrowLine(10,10)(50,45)
\ArrowLine(170,10)(130,45)
\ArrowLine(50,45)(90,45)
\ArrowLine(130,45)(90,45)
\DashArrowLine(10,80)(50,45){3}
\DashArrowLine(170,80)(130,45){3}
\Text(90,45)[]{$\mbox{\boldmath$\times$}$}
\Text(30,18)[]{$l$}
\Text(145,18)[]{$l$}
\Text(60,37)[]{$\nu^c$}
\Text(118,37)[]{$\nu^c$}
\Text(30,75)[]{$H$}
\Text(150,75)[]{$H$}
\Text(90,55)[]{$M_{\rm Maj}$}
\end{picture}
\end{center}
\caption{Diagram generating the dimension 5 operator (\ref{eqn:dim5op}).}
\label{fig:rge1}
\end{figure}
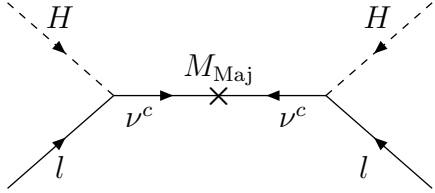

It is instructive to repeat the decoupling procedure in the field theoretical
language. The effects of the Born diagram,  Fig.~\ref{fig:rge1}, present
in the full theory valid above the scale $M$ are, up to ${\cal O}(1/M^2)$, 
reproduced in the effective theory, describing physics below the scale $M$, 
by a non-renormalizable operator of dimension 5 \cite{WEI}:
\begin{eqnarray}
\Delta{\cal L}=-{1\over4M}\mbox{\boldmath$C$}^{AB}
(\epsilon_{ij}H_il^A_j)(\epsilon_{lk}H_ll^B_k) + {\rm H.c.}
\label{eqn:dim5op}
\end{eqnarray}
in which
\begin{eqnarray}
M^{-1}\mbox{\boldmath$C$}=\mbox{\boldmath$Y$}_\nu^T
(\mbox{\boldmath$M$}_{\rm Maj})^{-1}\mbox{\boldmath$Y$}_\nu .\label{eqn:ymy}
\end{eqnarray}
The overall scale $M$ of the entries of the matrix 
$\mbox{\boldmath$M$}_{\rm Maj}$ has been factorized out from
\mbox{\boldmath$C$} to make it dimensionless. Thus, if the scale $M$ is
high enough, the effects of the Majorana masses can be discussed in
the SM or MSSM supplemented by the operator (\ref{eqn:dim5op}). We should
also stress that this operator is the only one of dimension 5 contributing
to the neutrino Majorana mass matrix. Other possible contributions are of
higher dimension \cite{BALE}. Thus, one may expect that in the effective SM
or MSSM the neutrino masses are indeed described by  the operator 
(\ref{eqn:dim5op}), even if its origin is different from the see-saw mechanism.
In particular, degenerate or partly degenerate patterns of neutrino masses
require some interplay between the parameters entering 
the formula (\ref{eqn:ymy}), but one can also imagine that such patterns 
originate from other mechanisms at the scale $M$. 

We end this section with the discussion of the parametrization of the neutrino 
mixing matrix. In the effective theory, after the electroweak symmetry 
breaking, the operator (\ref{eqn:dim5op}) is the source of the Majorana 
mass for the three active neutrinos $\nu$, ${\cal L}_{\rm mass}=-{1\over2}
{v^2\over4M}\mbox{\boldmath$C$}^{AB}\nu^A\nu^B+{\rm H.c.}$ This mass matrix 
can be diagonalized by an additional unitary rotation 
$\nu^A\rightarrow U^{Aa}\nu_a$. Recall we work in the basis in which the 
leptonic Yukawa coupling is already diagonal and the matrix
$\mbox{\boldmath$C$}$ is assumed to be given in the same basis. We have:
\begin{eqnarray}
{v^2\over4M}\left(U^T\mbox{\boldmath$C$}U\right)^{ab} = 
{v^2\over4M}C^a\delta^{ab} = m_{\nu_a}\delta^{ab}
\label{eqn:diagmass}
\end{eqnarray}
The matrix $U$ will therefore appear in the couplings of neutrinos to $W^\pm$ 
bosons and charged Goldstone (and Higgs) boson.\footnote{In 
a general electroweak basis in which the leptonic Yukawa coupling is not 
necessarily diagonal the matrix $U$ is given by the product 
$U=\mbox{\boldmath$E$}_L^\dagger\mbox{\boldmath$V$}_L$
where the rotations $e\rightarrow\mbox{\boldmath$E$}_Le$ and 
$\nu\rightarrow\mbox{\boldmath$V$}_L\nu$ diagonalize the mass matrices
of charged leptons and neutrinos, respectively.}
Being unitary, the matrix $U$ depends on 3 angles and 6 phases and can 
be conveniently written as $U^{Aa}=e^{i\varphi_A}U_{\rm MNS}^{Aa}$ where 
$U_{\rm MNS}$ is given in eq.~(\ref{eqn:MNSmat}). Contrary to the quark case,
if the Majorana masses $m_{\nu_a}$ in eq.~(\ref{eqn:diagmass}) are to be 
real and positive, the only freedom that remains is the possibility to 
re-phase independently the three Dirac fields of the charged leptons
$\psi_{e_A}\rightarrow\exp(i\varphi_A)\psi_{e_A}$ (i.e. 
$e_A\rightarrow\exp(i\varphi_A)e_A$ for the left-handed and 
$e^c_A\rightarrow\exp(-i\varphi_A)e^c_A$ for their right-handed components).
It is then the MNS matrix (\ref{eqn:MNSmat}) with three angles and three 
phases that enters the interactions of neutrinos with the SM particles. In 
the MSSM, however, the phases $\varphi_A$ can be eliminated (by appropriate 
rotations of the superpartner fields) only if the slepton mass matrices and 
charged lepton mass matrices are simultaneously diagonal. In the general 
case, the phases $\varphi_A$ appear in the neutrino-sneutrino-neutralino and 
neutrino-chargino-charged slepton vertices.

It is sometimes convenient to work with complex neutrino masses
$m_{\nu_a}$ in eq.~(\ref{eqn:diagmass}). The phases $\alpha_{1,2}$ from 
eq.~(\ref{eqn:MNSmat}) are then absorbed into the masses:
$m_{\nu_{1,2}}=|m_{\nu_{1,2}}|\exp(-2i\alpha_{1,2})$. In particular, when
CP is conserved the phases $\alpha_{1,2}=0,\pm\pi/2$ reflecting different 
CP parities of different neutrinos, can be absorbed into the mass eigenvalues 
and make them real positive or negative. The MNS matrix is then real. Since 
only relative signs of the masses play a role, we will for definiteness fix 
our convention so that $m_{\nu_3}$ is always positive. Such a parametrization 
simplifies greatly the qualitative analysis of the renormalization group 
equations discussed in the next section.

The value of the mass $M$ is not known. One can expect that it is determined 
by a beyond the SM theory that provides the physical cut-off for the SM. The 
idea of Grand Unification and of the  big desert would suggest very high value 
of $M$. However there are also other models (large extra dimensions) in which 
$M$ is ${\cal O}(1$ TeV). In this review we assume that whatever the theory of 
neutrino masses is, the scale $M$ is high enough to justify the description of 
the neutrino mass effects by the effective theory (SM or MSSM) with the single
operator (\ref{eqn:dim5op}) added.

Finally, one should also mention that in the MSSM neutrino mass can originate 
{} from R-parity violating interactions \cite{HE}. Such models are usually
based on low energy mechanisms for neutrino mass generation and will not be
discussed in this review.

\section{Quantum corrections from the renormalization group evolution}
\setcounter{equation}{0} 

The neutrino mass parameters given at some scale $M$ in the effective theory 
(SM or MSSM) supplemented by the operator (\ref{eqn:dim5op}), determine the 
measurable quantities after inclusion of quantum corrections. The first 
category  of corrections are those described by the evolution of the effective
theory parameters from the scale $M$ down to some low energy scale close to 
the electroweak scale. They depend on arbitrary powers of the large logarithms
of the ratio $M/M_Z$  and are resummed to all orders of the perturbation 
expansion by means 
of the renormalization group equations (RGE). The second class of corrections 
are the so-called low energy threshold corrections. In the SM they connect the 
mass parameters of the SM Lagrangian renormalized at the scale $M_Z$ to the 
fermion masses in the the effective theory obtained after the electroweak 
symmetry breaking (whose renormalizable part is QCD+QED) renormalized at the 
same scale $M_Z$. In the MSSM, threshold corrections include also the 
superpartner contributions due to their mass splittings and/or 
flavour violation in slepton mass matrices. The relative
magnitude of the two types of corrections will be discussed later on.
In this section we assume that the RG corrections are the dominant ones and,
after deriving the relevant RG equations, we discuss the potential role
played by these corrections for the neutrino masses and mixing.

\subsection{RG equations for the CKM matrix}

Before we derive the RGE for the neutrino masses and mixing we discuss the 
technically easier case of the evolution of the Yukawa matrices and the CKM 
matrix.

The renormalization group equations for the Yukawa couplings of the SM and
the MSSM at one \cite{CHEILI,MAVA} and two loop \cite{MAVA,BABEOH1} are well 
known. For convenience we reproduce the one loop RGE in Appendix B. They 
have a matrix structure which encodes the evolution of their eigenvalues 
and of all entries of the CKM mixing matrix. From the full RGE in the matrix 
form it is possible \cite{BABU,OLPO} to derive the RGE for the eigenvalues of 
the Yukawa matrices and for the CKM matrix. It is often very instructive to 
discuss the evolution of 
those Lagrangian parameters that can be directly determined from the data. 
For instance, the experimentally known information can be unambiguously 
extrapolated to any high scale. Moreover, certain qualitative features of the 
Yukawa coupling pattern necessary to reproduce the experimental data can be 
easier to understand. The derivation goes as follows. One writes down 
auxiliary RGEs for the matrices $\mbox{\boldmath$U$}_{L,R}$, 
$\mbox{\boldmath$D$}_{L,R}$ defined in eq.~(\ref{eqn:rotations}) in the form:
\begin{eqnarray}
{d\over dt}\mbox{\boldmath$U$}_L=\mbox{\boldmath$U$}_L\varepsilon^U_L, ~~~
{d\over dt}\mbox{\boldmath$U$}_R=\mbox{\boldmath$U$}_R\varepsilon^U_R,
\nonumber\\
{d\over dt}\mbox{\boldmath$D$}_L=\mbox{\boldmath$D$}_L\varepsilon^D_L, ~~~
{d\over dt}\mbox{\boldmath$D$}_R=\mbox{\boldmath$D$}_R\varepsilon^D_R,
\label{eqn:epsmatrices}
\end{eqnarray}
where the matrices $\varepsilon^{U,D}_{L,R}$ are antihermitean, in order 
to preserve unitarity of the matrices $\mbox{\boldmath$D$}_{L,R}$,
and $\mbox{\boldmath$U$}_{L,R}$, and $t=(1/16\pi^2)\ln(Q/M_Z)$. One then looks
for the evolution of the $\mbox{\boldmath$U$}_{L,R}$ and 
$\mbox{\boldmath$D$}_{L,R}$ matrices such that Yukawa matrices remain
diagonal during the evolution.
Differentiating the four relations (obtained from eqs.~(\ref{eqn:diagyuk})) 
\begin{eqnarray}
{\rm diag}(y^2_u, ~y^2_c, ~y^2_t)
=\mbox{\boldmath$U$}_L^\dagger\mbox{\boldmath$Y$}_u^\dagger
\mbox{\boldmath$Y$}_u\mbox{\boldmath$U$}_L
=\mbox{\boldmath$U$}_R^\dagger\mbox{\boldmath$Y$}_u
\mbox{\boldmath$Y$}_u^\dagger\mbox{\boldmath$U$}_R\nonumber 
\end{eqnarray}
\begin{eqnarray}
{\rm diag}(y^2_d, ~y^2_s, ~y^2_b)
=\mbox{\boldmath$D$}_L^\dagger\mbox{\boldmath$Y$}_d^\dagger
\mbox{\boldmath$Y$}_d\mbox{\boldmath$D$}_L
=\mbox{\boldmath$D$}_R^\dagger\mbox{\boldmath$Y$}_d
\mbox{\boldmath$Y$}_d^\dagger\mbox{\boldmath$D$}_R
\end{eqnarray}
and requiring the derivatives of the diagonal matrices on the rhs to
be also diagonal matrices one gets the matrices $\varepsilon^{U,D}_L$:
\begin{eqnarray}
(\varepsilon^U_L)_{JI}=-u_d{y^2_{u_J}+y^2_{u_I}\over y^2_{u_J}-y^2_{u_I}}
\sum_KV^{JK}V^{IK\ast}y_{d_K}^2, ~~~
(\varepsilon^U_L)_{JJ}=0\nonumber\\
(\varepsilon^D_L)_{KL}=-d_u{y^2_{d_K}+y^2_{d_L}\over y^2_{d_K}-y^2_{d_L}}
\sum_Jy^2_{u_J}V^{JK\ast}V^{JL}, ~~~
(\varepsilon^D_L)_{KK}=0
\label{eqn:epsqL}
\end{eqnarray}
and similarly the matrices $\varepsilon^{U,D}_R$:
\begin{eqnarray}
(\varepsilon^U_R)_{JI}=-u_d{2y_{u_J}y_{u_I}\over y^2_{u_J}-y^2_{u_I}}
\sum_KV^{JK}V^{IK\ast}y_{d_K}^2, ~~~
(\varepsilon^U_R)_{JJ}=0\nonumber\\
(\varepsilon^D_R)_{KL}=-d_u{2y_{d_K}y_{d_L}\over y^2_{d_K}-y^2_{d_L}}
\sum_Jy^2_{u_J}V^{JK\ast}V^{JL}, ~~~
(\varepsilon^D_R)_{KK}=0
\end{eqnarray}
where $u_d=d_u=-3/2$ for the SM and $u_d=d_u=1$ for the MSSM. The evolution 
of the CKM mixing matrix $V$ is then given by the simple matrix equation
\begin{eqnarray}
{d\over dt}V=-\varepsilon^U_LV+V\varepsilon^D_L.
\label{eqn:rgeckm}
\end{eqnarray}
Because of the hierarchical pattern of quark masses the expressions 
(\ref{eqn:epsqL}) for $\varepsilon^{U,D}_L$ can be simplified to
\begin{eqnarray}
(\varepsilon^D_L)_{KL}=-d_uy^2_tV^{3K\ast}V^{3L} ~~{\rm for} ~~K>L &&
{\rm and}  ~~~(\varepsilon^D_L)_{KL}=-(\varepsilon^D_L)_{LK} ~~{\rm for} ~~K<L
\nonumber\\
(\varepsilon^U_L)_{JI}=-u_dV^{J3}V^{I3\ast}y_b^2 ~~{\rm for} ~~J>I &&
{\rm and}  ~~~(\varepsilon^U_L)_{JI}=-(\varepsilon^U_L)_{IJ} ~~{\rm for} ~~J<I.
\label{eqn:epsqLsimpl}
\end{eqnarray}
Taking next into account the hierarchy of the entries of the CKM matrix it is 
easy to find that, to a good approximation, the $2\times2$ submatrix describing
the mixing of the first two generations as well as the element $V^{33}$ 
do not evolve, while the evolution of the remaining entries is universal and 
is given by \cite{OLPO,BABEOH2}:
\begin{eqnarray}
V(t)=V(0)\exp\left\{-\int_0^t\left[u_dy^2_b(t^\prime)
+d_uy^2_t(t^\prime)\right]dt^\prime\right\}
\end{eqnarray}
where $V=V^{31}$, $V^{32}$, $V^{13}$ or $V^{32}$. It follows that to one 
loop the Jarlskog
invariant $J\equiv{\rm Im}(V^{ud}V^{cs}V^{us\ast}V^{cd\ast})$ does not change.
By using e.g. the standard parametrization of the CKM matrix it is also easy 
to derive the RGE for the three mixing angles and the CP violating 
phase.\footnote{In the one loop approximation the running of the CKM matrix 
elements looks particularly simple in the Wolfenstein parametrization 
\cite{WO} (see e.g. \cite{POKBOOK}) in which only the parameter $A$ 
changes with the scale.} 
For the SM and the MSSM this has been done up to two-loops in 
ref.~\cite{BABEOH2}.

\subsection{RGEs for neutrino masses and mixing}

The Majorana mass term for neutrinos arises from the dimension 5 operator
(\ref{eqn:dim5op}). The RGE for $\mbox{\boldmath$C$}^{AB}$ in the SM has been 
correctly computed only recently in ref.~\cite{ANDRKELIRA} (previous 
calculations \cite{WE,CHPL,BALEPA} have errors) and in the MSSM in 
refs.~\cite{CHPL,BALEPA}:\footnote{The MSSM RGEs given in ref.~\cite{CHPL}
allow to treat also the case in which squarks and/or gluino are much heavier
than sleptons, charginos and neutralinos so that the decoupling procedure
\cite{CH} can be employed; there are then four different operators which mix 
with each other below the squark/gluino threshold. Above it one has
(in the notation of ref.~\cite{CHPL}) 
$c_1^{ab}=2c_{12}^{ab}=2c_{21}^{ab}=c_3^{ab}\equiv\mbox{\boldmath$C$}^{ab}$
and the four equations of ref.~\cite{CHPL} merge into the one quoted here.}
\begin{eqnarray}
{d\over dt}\mbox{\boldmath$C$} = -K\mbox{\boldmath$C$} 
-\kappa\left[\left(\mbox{\boldmath$Y$}_e^\dagger\mbox{\boldmath$Y$}_e\right)^T
\mbox{\boldmath$C$} + \mbox{\boldmath$C$}
\left(\mbox{\boldmath$Y$}_e^\dagger\mbox{\boldmath$Y$}_e\right)\right]
\label{eqn:rgeCgen}
\end{eqnarray}
where in the SM $\kappa=-3/2$ and $K=-3g^2_2
+2{\rm Tr}\left(3\mbox{\boldmath$Y$}_u^\dagger\mbox{\boldmath$Y$}_u
+3\mbox{\boldmath$Y$}_d^\dagger\mbox{\boldmath$Y$}_d
+\mbox{\boldmath$Y$}_e^\dagger\mbox{\boldmath$Y$}_e\right)
+2\lambda$; the normalization of $\lambda$ is fixed by the Higgs self
interaction: ${\cal L}_{\rm self}=-{\lambda\over2}(H^\dagger H)^2$. 
In the MSSM, $\kappa=+1$ and $K=-6g^2_2 - 2g^2_Y
+2{\rm Tr}\left(3\mbox{\boldmath$Y$}_u^\dagger\mbox{\boldmath$Y$}_u\right)$.
Since our initial conditions will be always given at the scale $M$, we have
written eq.~(\ref{eqn:rgeCgen}) top-down, i.e. with $16\pi^2t=\ln(M/Q)$.
The equation (\ref{eqn:rgeCgen}) is valid in any flavour basis, in particular 
in the basis in which the Yukawa matrix $\mbox{\boldmath$Y$}_e$ is 
diagonal.\footnote{This is because the matrix $\mbox{\boldmath$E$}_L$ 
defined in eq.~(\ref{eqn:rotations}) does not evolve; the matrices 
$\varepsilon_{L,R}^E$ analogous to the ones defined in 
eq.~(\ref{eqn:epsmatrices}) are identically zero because below the scale 
$M$, at which the operator (\ref{eqn:dim5op}) is generated, the RGE for the 
leptonic Yukawa coupling (\ref{eqn:dpoptY_sm},\ref{eqn:dpoptY_mssm}) do not
depend on the couplings $\mbox{\boldmath$Y$}_\nu$, the only one that could 
change their matrix structure during the evolution.}
In that basis, eq.~(\ref{eqn:rgeCgen}) simplifies to
\begin{eqnarray}
{d\over dt}\mbox{\boldmath$C$}^{AB}=-K\mbox{\boldmath$C$} ^{AB}
-\kappa\left[y^2_{e_A}\mbox{\boldmath$C$}^{AB} + \mbox{\boldmath$C$}^{AB}
y^2_{e_B}\right]
\end{eqnarray}
where $y^2_{e_A}$ are the eigenvalues of the hermitean matrix 
$\mbox{\boldmath$Y$}_e^\dagger\mbox{\boldmath$Y$}_e$.
In this form it can be elegantly solved \cite{ELLO}:
\begin{eqnarray}
\mbox{\boldmath$C$}(t)= I_K{\cal J}\mbox{\boldmath$C$}(0){\cal J}
\label{eqn:ellosol}
\end{eqnarray}
where ${\cal J}={\rm diag}(I_e,I_\mu,I_\tau)$ and
\begin{eqnarray}
I_K=\exp\left(-\int_0^tK(t^\prime)dt^\prime\right),\nonumber\\
I_{e_A}=\exp\left(-\kappa\int_0^ty^2_{e_A}(t^\prime)dt^\prime\right).
\label{eqn:lolaIfact}
\end{eqnarray}
Note that any zero in the initial matrix $\mbox{\boldmath$C$}^{AB}$ is 
preserved by the RG evolution and that the phases of 
the $\mbox{\boldmath$C$}^{AB}$ entries do not evolve \cite{HAMAOKSU1}.
We also note that, since $I_KI_e^2\approx1$, the experimental constraint 
(\ref{eqn:2bb0nubound}), with 
$m_\nu^{ee}=(v^2/4M)\mbox{\boldmath$C$}^{11}(0)$ and $v^2=v^2(M_Z)$, 
is renormalized negligibly.

Although the solution (\ref{eqn:ellosol}) to the RG equation for 
$\mbox{\boldmath$C$}^{AB}$ is simple, qualitative features of the running
of the neutrino mass eigenvalues $m_{\nu_a}$ 
and of the MNS matrix are often masked by the diagonalization procedure.
In that approach it has to be performed {\it after} the evolution. It is
therefore useful to derive the RGE directly for the mass eigenvalues and
mixing angles. This is done by using a trick similar as for
the CKM matrix. One defines the auxiliary antihermitean matrix 
$\varepsilon^\nu$  by the equation
\begin{eqnarray}
{d\over dt}U = U\varepsilon^\nu
\label{eqn:ddtU}
\end{eqnarray}
where $U$ satisfies eq.~(\ref{eqn:diagmass}). Differentiating the equality
$C^a\delta^{ab}=(U^T\mbox{\boldmath$C$}U)^{ab}$ and requiring that its rhs 
remains a diagonal matrix gives us the matrix $\varepsilon^\nu$. The only 
difference is that the resulting equation depends on $(\varepsilon^\nu)^T$ 
instead of $(\varepsilon^\nu)^\dagger$. Hence, one obtains two separate 
equations for ${\rm Re}(\varepsilon^\nu)$ and ${\rm Im}(\varepsilon^\nu)$:
\begin{eqnarray}
&&{\rm Re}(\varepsilon^\nu)^{ab}=\kappa A_{ab}
{\rm Re}\left(\sum_AU^{Aa\ast}y^2_{e_A}U^{Ab}\right)\nonumber\\
&&{\rm Im}(\varepsilon^\nu)^{ab}=\kappa(A_{ab})^{-1}
{\rm Im}\left(\sum_AU^{Aa\ast}y^2_{e_A}U^{Ab}\right)
\label{eqn:eps_nu}
\end{eqnarray}
where 
\begin{eqnarray}
A_{ab}\equiv{C^a+C^b\over C^a-C^b}=
{m_{\nu_a}+m_{\nu_b}\over m_{\nu_a}-m_{\nu_b}}
\label{eqn:Aab}
\end{eqnarray}
and of course ${\rm Re}(\varepsilon^\nu)^{aa}=
{\rm Im}(\varepsilon^\nu)^{aa}=0$. For conserved CP ($U$ real), those 
equations have been first derived in ref.~\cite{CHKRPO}. For a general 
complex matrix $U$ they have been given in \cite{CAESIBNA4}. The formulae 
for $dU/dt$ written there is valid in a general electroweak basis; it reduces 
to (\ref{eqn:eps_nu}) after setting 
$P=\mbox{\boldmath$Y$}_e^\dagger\mbox{\boldmath$Y$}_e=P^\dagger$ and 
passing to the basis in which $P$ is diagonal. Thus,
\begin{eqnarray}
&&{d\over dt}U^{Aa} = \kappa\sum_{b\neq a}U^{Ab}\left[A_{ba}
{\rm Re}\left(\sum_BU^{Bb\ast}y^2_{e_B}U^{Ba}\right) + {i\over A_{ba}}
{\rm Im}\left(\sum_BU^{Bb\ast}y^2_{e_B}U^{Ba}\right)\right]\phantom{aaaa}
\label{eqn:Urun}\\
&&{d\over dt}C^a = -\left(K+ 2\sum_Ay^2_{e_A}|U^{Aa}|^2\right)C^a.
\label{eqn:Crun}
\end{eqnarray}

Eq.~(\ref{eqn:Urun}) gives directly the running of the angles $\theta_{12}$, 
$\theta_{23}$, $\theta_{13}$ \cite{CHKRPO} and the phases $\delta$, 
$\alpha_a$, $\varphi_A$ parameterizing the matrix 
$U^{Aa}=e^{i\varphi_A}U^{Aa}_{\rm MNS}$ \cite{CAESIBNA4}. In general, 
even if the phases $\varphi_A$ are zero at some scale they will be 
generated during the evolution. However, it is easy to see that the 
differential equations for the parameters of the MNS matrix 
(\ref{eqn:MNSmat}) do not depend on the $\varphi_A$.
Indeed, from eq.~(\ref{eqn:eps_nu}) 
it follows that $\varepsilon^\nu$ does not depend on $\varphi_A$; 
furthermore, since $\dot U^{Aa}=e^{i\varphi_A}[\dot U^{Aa}_{\rm MNS} + 
i\dot\varphi_AU^{Aa}_{\rm MNS}]$, all factors $e^{i\varphi_A}$ cancel 
out in eq.~(\ref{eqn:ddtU}).
 
If two of  the three eigenvalues, say  $m_{\nu_a}$  and $m_{\nu_b}$, are 
equal at some  scale  $t$ there is the  freedom  in choosing the matrix 
$U(t)$, corresponding  to the redefinition $U(t)\rightarrow\tilde U(t)=U(t)R$
where $R$ is a rotation in the $ab$ plane. For  the evolution, $R$  has  
to  be fixed  by the condition 
\begin{eqnarray}
{\rm Re}[\sum_A\tilde U^{Aa\ast} y^2_{e_A}\tilde U^{Ab}](t) =0 
\label{eqn:crosscond}
\end{eqnarray}
so that $\varepsilon^\nu$ in nonsingular. This is particularly important for 
considering mass patterns with exact degeneracy at the scale $M$. The mixing 
matrix $U$ is then ambiguous at that scale in the tree level approximation
and becomes determined by quantum corrections, no matter how small. We shall 
illustrate this point in the next section. Note also that for conserved CP 
and neutrinos $\nu_a$ and $\nu_b$ of opposite CP parities such an ambiguity 
of rotations $R$ does not exist. In such a case it is convenient to work with 
a real matrix $U^{Aa}$ and with the neutrino masses of opposite signs (see the
comments at the end of Section 3). The RG equation (\ref{eqn:Urun}) is then 
nonsingular because $A_{ab}$, instead of being divergent, vanishes.  

Eqs.~(\ref{eqn:Urun}) and (\ref{eqn:Crun}) are very convenient for a
qualitative discussion of the impact of the RG evolution directly on the
neutrino masses and mixing angles. As we shall see, several physical
effects are in this approach more transparent than in the approach based on
evolving the matrix $\mbox{\boldmath$C$}^{AB}$. 

\subsection{Evolution of the neutrino masses}

The effects of the RG quantum corrections on the neutrino mass eigenvalues
are simple and, except for a few special cases, not very interesting. The
solution to eq.~(\ref{eqn:Crun}) reads
\begin{eqnarray}
m_{\nu_a}(t) = I_K\exp\left(-2\kappa\int_0^ty^2_{e_A}(t^\prime)
|U^{Aa}(t^\prime)|^2dt^\prime\right) m_{\nu_a}(0).
\label{eqn:massrge}
\end{eqnarray}
Neglecting the small $y_e$ and  $y_\mu$ Yukawa couplings\footnote{In the SM 
the Yukawa couplings are unambiguously determined at the electroweak scale by 
the charged lepton masses; in the MSSM 
$(y^2_{e_A})_{\rm MSSM}=(y^2_{e_A})_{\rm SM}/\cos\beta$ where 
$\tan\beta\equiv v_u/v_d$ can vary from a few up to $\sim50$.}
\begin{eqnarray}
m^2_{\nu_a}(t)=m^2_{\nu_a}(0)I^2_K\exp\left(-4\kappa\int_0^ty^2_\tau(t^\prime)
|U_{3a}|^2(t^\prime)dt^\prime\right) 
\label{eqn:solmass}
\end{eqnarray}
Factors $I_K$ and $\kappa$ are different for the SM and the MSSM
($\kappa=-3/2$ and $+1$, respectively). We see that the possibility of some  
change in the mass pattern caused by the evolution resides solely in the 
differences in the mixing matrix elements $U_{3a}$ and their RG running.   
In the MSSM where $y^2_\tau\approx(\tan\beta/100)^2$, $t_Z\approx0.12$ for 
$M=10^{10}$ GeV and $U^2_{3a}$ typically varies between $0$ and $1/4$ 
(except for $U^2_{33}$), the exponent is at most of order of $\epsilon\equiv
y^2_\tau\log(M/M_Z)/16\pi^2\approx\tan^2\beta\times10^{-5}<2.5\times10^{-2}$ 
for $\tan\beta<50$. In the SM, $y^2_\tau\approx10^{-4}$ and, consequently,
$\epsilon\approx10^{-5}$. We can then estimate the changes
in the mass squared differences: 
\begin{eqnarray}
\Delta m^2_{ab}(t)\equiv m^2_{\nu_a}(t)-m^2_{\nu_b}(t)=\Delta m^2_{ab}(0)    
- (\eta_a m^2_{\nu_a}(0)-\eta_b m^2_{\nu_b}(0))\epsilon 
\label{eqn:runmassdiff}
\end{eqnarray}
where we have neglected $I_K$ which is always close to 1, and the factors 
$\eta_{a(b)}>0$ are typically in the range $0-2$, depending on the values 
of $U_{3a}$ factors and their evolution. Taking (for definiteness) 
$\Delta m^2_{ab}(0)=0$, we see that the evolution of $\Delta m^2_{ab}(t)$ 
is limited by $m^2_{\nu_a}(0)\epsilon$ or $m^2_{\nu_b}(0)\epsilon$, i.e. by 
the value of the larger mass. 

We see that the RG evolution cannot change the pattern of masses. It may still
be of importance for precision tests of various models, particularly those 
with partly degenerate or degenerate patterns. We recall here that such 
patterns have (at least approximately) degenerate two and three neutrinos,
respectively, at the scale $M$. Matching the experimental $\Delta m^2$ and
$\Delta M^2$ needs then some fine-tuning between the initial values and the
RG quantum corrections. That point can, however, be meaningfully discussed
only for each concrete model.

The special cases are those with equal masses, 
$m^2_{\nu_1}=m^2_{\nu_2}\neq m^2_{\nu_3}$ or 
$m^2_{\nu_1}=m^2_{\nu_2}= m^2_{\nu_3}$ at the scale $M$. They are called in
this review two-fold and three-fold degeneracies, respectively, to be 
distinguishable from the more general partly degenerate or degenerate 
patterns. The interesting question we shall discuss in Subsection 4.5 and
Section 5 is whether only quantum corrections can then explain the observed
mass squared differences.

In the next sections we shall discuss the RG quantum corrections to the 
neutrino mixing. We shall focus on interesting qualitative effects that
depend only on the broad classification of the neutrino mass pattern, and 
do not depend on such details as whether the masses are equal or only 
approximately equal. 

\subsection{Mixing of two neutrinos}

We first examine the mixing of two neutrinos, which was investigated in many 
papers \cite{BALEPA,TA,HAOKSU,BADIMOPA1,BAMOPAPA}. Strictly speaking, it 
could be physically relevant only if the atmospheric neutrino anomaly was 
due to the $\nu_\mu$-$\nu_\tau$ oscillations whereas the solar neutrino 
deficit resulted from the $\nu_e$-$\nu_{\rm sterile}$ oscillations (at 
present strongly disfavoured by the SNO and Superkamiokande data). 
Moreover the two $2\times2$ neutrino systems would have to be 
completely independent due to some particular texture of the $U$ matrix. 
Nevertheless we will see, that the evolution of the $\theta_{23}$ angle
in the $3\times3$ scenario is, in some cases very similar to the evolution 
of the mixing angle of two neutrinos only. It is therefore
instructive to discuss the $2\times2$ evolution and to
compare it later with the more realistic $3\times3$ mixing.

We begin with a real matrix $\mbox{\boldmath$C$}^{AB}$ and hence, a 
real $U^{Aa}$:
\begin{eqnarray}
U^{Aa}=\left(\matrix{c_\vartheta&-s_\vartheta\cr 
s_\vartheta&c_\vartheta}\right)
\end{eqnarray}
where $s_\vartheta\equiv\sin\vartheta$, $c_\vartheta\equiv\cos\vartheta$. 
We shall consider the same or different CP parities of the two neutrinos
$m_{\nu_1}m_{\nu_2}>0$ or $m_{\nu_1}m_{\nu_2}<0$, respectively.  From 
eq.~(\ref{eqn:Urun}) one finds the following equations
for the mixing angle
\begin{eqnarray}
{d s_\vartheta\over dt}=\kappa A_{21}(y^2_{e_2}-y^2_{e_1})
s_\vartheta c^2_\vartheta ~~~
{d c_\vartheta\over dt}=-\kappa A_{21}(y^2_{e_2}-y^2_{e_1})
s^2_\vartheta c_\vartheta,
\label{eqn:ddt2x2}
\end{eqnarray}
where $\kappa=-3/2$ and $+1$ in the SM and the MSSM, respectively. 
Eqs.~(\ref{eqn:ddt2x2}) give
\begin{eqnarray}
{d\over dt}\sin^22\vartheta=2\kappa A_{21}\left(y^2_{e_2}-y^2_{e_1}\right)
\sin^22\vartheta\cos2\vartheta
\label{eqn:ddtsin22}
\end{eqnarray}
known from the literature\footnote{Its rhs is more frequently written in the
equivalent form $2\kappa{C^{22}+C^{11}\over C^{22}-C^{11}}
\left(y^2_{e_2}-y^2_{e_1}\right)\sin^22\vartheta\cos^22\vartheta$.}
\cite{BALEPA,TA,ELLELONA,LO,CAELLOWA}. Eq.~(\ref{eqn:ddtsin22}) has a trivial 
fixed point (FP) at $\sin^22\vartheta=0$ but, contrary to the statements made 
in some papers, the maximal mixing $\sin^22\vartheta=1$ is not 
its FP: Although, naively, $\sin^22\vartheta(t)\equiv1$ solves 
eq.~(\ref{eqn:ddtsin22}), it is easy to see that $s_\vartheta(t)=\pm1/\sqrt2$ 
does not solve eqs.~(\ref{eqn:ddt2x2}).\footnote{For example, for constant
$\eta\equiv\kappa A_{21}(y^2_{e_2}-y^2_{e_1})$, eq.~(\ref{eqn:ddt2x2}) is 
solved by $s^2_\vartheta(t)=s^2_0/(s^2_0+c^2_0\xi)$ where 
$\xi\equiv\exp(-2\eta t)$ and $s_0$ is the initial value of $s_\vartheta$.
It is then straightforward to check that 
$\sin^22\vartheta(t)=\sin^22\vartheta_0\xi/(s^2_0+c^2_0\xi)^2$ solves 
eq.~(\ref{eqn:ddtsin22}) for any initial value $\sin^22\vartheta_0$. Thus, 
for  $\sin^22\vartheta_0=1$, eq.~(\ref{eqn:ddtsin22}) has two solutions: 
$\sin^22\vartheta(t)\equiv1$ and the one given here but only the latter 
satisfies the underlying eq.~(\ref{eqn:ddt2x2}).}

It is clear from eq.~(\ref{eqn:ddt2x2}) that, for fixed values of the Yukawa
couplings $y^2_{e_A}$ and for some fixed evolution ``time'' 
$t_Z\equiv(1/16\pi^2)\ln(M/M_Z)$, the evolution of the mixing 
angle depends on the factor $A_{21}=(m_{\nu_2}+m_{\nu_1})^2/\Delta m^2_{21}$.
It is always
small when the two neutrinos have opposite CP-parities i.e. when 
$m_{\nu_1}m_{\nu_2}<0$ (as follows from eq.~(\ref{eqn:massrge}), 
$m_{\nu_a}$ cannot change sign during the evolution) or if the neutrino mass
spectrum is hierarchical ($m^2_{\nu_1}\ll m^2_{\nu_2}$ or 
$m^2_{\nu_2}\ll m^2_{\nu_1}$). The evolution
of the mixing angle is then negligible \cite{HAOKSU,HAMAOKSU1,MA,ELLO}.
For $m_{\nu_1}m_{\nu_2}>0$ the evolution of $\vartheta$ can be significant,
particularly if the neutrino masses are nearly degenerate so that 
$|A_{21}\epsilon|\gg1$ where $\epsilon\equiv(y^2_{e_2}-y^2_{e_1})t_Z$.
It can be checked that for $|A_{21}|\epsilon\simgt3$ the FP at 
$\sin^22\vartheta=0$ is reached at the electroweak scale.

In view of the maximal $\nu_\mu$-$\nu_\tau$ mixing needed to explain the 
atmospheric neutrino data, some attention was paid to the possibility of 
increasing $\vartheta$ by the RG corrections from a small value at the $M$ 
scale to a (nearly) maximal ($|\vartheta|\approx\pi/4$) at the electroweak
scale \cite{BALEPA,TA,HAOKSU,BADIMOPA1,BAMOPAPA}. As we said earlier, the 
maximal mixing is not a FP of eqs.~(\ref{eqn:ddt2x2}), so such an 
``explanation'' of large mixing can merely be due to a coincidence of the 
running ``time'' and the initial values of the angles and the masses. The 
value of $|A_{21}|\epsilon$ must be in the range such that the evolution is
non-negligible, but not strong enough to reach the FP. This 
can be most easily seen if we return to the solution (\ref{eqn:ellosol}) 
written in the form
\begin{eqnarray}
\mbox{\boldmath$C$}^{AB}(t)=
\left(\matrix{C^{11}(t)&C^{12}(t)\cr C^{12}(t)&C^{22}(t)}\right)
\propto\left(\matrix{C^{11}_0&C^{12}_0I_{21}(t)\cr 
C^{12}_0I_{21}(t)&C^{22}_0I_{21}^2(t)}\right)
\end{eqnarray}
where $I_{21}(t)=I_{e_2}/I_{e_1}=\exp\left(-\kappa\int_0^t(y^2_{e_2}-y^2_{e_1})
(t^\prime)dt^\prime\right)$. One then has
\begin{eqnarray}
\sin^22\vartheta(t)={4\left[C^{12}(t)\right]^2\over
\left[C^{11}(t)-C^{22}(t)\right]^2 + 4\left[C^{12}(t)\right]^2}
\end{eqnarray}

It is obvious that $\sin^22\vartheta(t)\approx1$ is obtained whenever 
$C^{11}(t)\approx C^{22}(t)$. Since $C^{22}(t)$ evolves differently from
$C^{11}(t)$ (as $y^2_{e_2}\gg y^2_{e_1}$) it is relatively easy to device 
the situation in which, at the initial scale $M$, $\sin^22\vartheta_0$ is 
small (this requires $|C^{11}_0-C^{22}_0|\gg 2|C^{12}_0|$) and the evolution 
is such that at some lower scale $C^{11}(t)= C^{22}(t)$ holds (this
obviously requires $C^{22}_0C^{11}_0>0$). With the judicious choice of the 
$\mbox{\boldmath$C$}^{AB}(0)$ matrix elements and 
$\epsilon\equiv(y^2_{e_2}-y^2_{e_1})t_Z$ it is possible
to obtain $\sin^22\vartheta\approx1$ at the electroweak scale
\cite{BADIMOPA1,BAMOPAPA}. Expressing $C^{11}_0$ and $C^{22}_0$ in terms of 
the neutrino masses and mixing angle at the initial scale $M$, the relevant 
condition $C^{11}_0=C^{22}_0I_{21}^2(t_Z)$ reads \cite{BAMOPAPA}
\begin{eqnarray}
(m_{\nu_1}^{(0)}c^2_{\vartheta_0}+m_{\nu_2}^{(0)}s^2_{\vartheta_0})=
(m_{\nu_1}^{(0)}s^2_{\vartheta_0}+m_{\nu_2}^{(0)}c^2_{\vartheta_0})
I^2_{21}(t_Z)\label{eqn:mohcond}
\end{eqnarray}
where $s_{\vartheta_0}\equiv\sin\vartheta(0)$ and 
$m_{\nu_a}^{(0)}\propto C^a(0)$.
It is clear that for $|m^{(0)}_{\nu_2}|\approx |m^{(0)}_{\nu_1}|$ and
$s_{\vartheta_0}\approx0$ (or $c_{\vartheta_0}\approx0$) satisfying the
condition (\ref{eqn:mohcond}) requires $m_{\nu_1}m_{\nu_2}>0$. On the other 
hand, for 
$m_{\nu_1}m_{\nu_2}<0$ the product $C^{11}_0C^{22}_0$ can be positive 
only if $m^2_{\nu_2}\ll m^2_{\nu_1}$ or $m^2_{\nu_2}\gg m^2_{\nu_1}$, which 
leads to a hierarchy $C^{11}_0\gg C^{22}_0$ or $C^{11}_0\ll C^{22}_0$. 
Getting $C^{22}(t)=C^{11}(t)$ requires then $I_{21}\ll1$ or $\gg1$. Thus, 
for $m_{\nu_1}m_{\nu_2}<0$ the evolution parameter $\epsilon$ must be 
large, too large to be accommodated in realistic theories. This is in agreement
with our earlier observation that for opposite CP parities of the two neutrinos
and/or their hierarchical masses the RG evolution is very weak.

Of course, since $\sin^22\vartheta=1$ is not the FP
of the RGE, $\sin^22\vartheta=1$ can hold only at one particular scale.

Similar strategy can be also applied to analyze the general  complex
$\mbox{\boldmath$C$}^{AB}$. Once the parameters of the $U^{Aa}$ matrix
\begin{eqnarray}
U=\left(\matrix{e^{i\varphi_1}&0\cr0&e^{i\varphi_2}}\right)
\left(\matrix{c_{\vartheta}&-s_{\vartheta}\cr 
      s_{\vartheta}&c_{\vartheta}}\right)
\left(\matrix{1&0\cr0&e^{i\alpha_2}}\right)
\end{eqnarray}
are expressed explicitly in terms of the $\mbox{\boldmath$C$}^{AB}$ entries, 
the solution (\ref{eqn:ellosol}) allows to obtain analytic formula for 
$\sin^22\theta(t)$ and to study its behaviour. The relevant formulae have 
been given in ref.~\cite{HAMAOK} (see also \cite{KUPAWU}). As could be 
expected, for fixed initial values of $\sin^22\theta$ and the neutrino masses 
$|m_{\nu_1}|$ and $|m_{\nu_2}|$, the $\sin^22\theta$ obtained by the RG 
evolution interpolates smoothly between its value obtained for $\alpha_2=0$ 
(i.e. $m_{\nu_1}m_{\nu_2}>0$) and 
$\alpha_2=\pm\pi/2$ (i.e. with $m_{\nu_1}m_{\nu_2}<0$) \cite{HAMAOKSU2}.

\subsection{Mixing of three neutrinos and fixed points}

For conserved CP, it is straightforward to derive from 
(\ref{eqn:Urun}) the equations  for  the three  independent mixing parameters 
$s_{12}$, $s_{23}$ and $s_{13}$ \cite{CHKRPO,CAESIBNA4}. Neglecting $y_e$ and 
$y_\mu$ Yukawa couplings we get, both in the SM and in the MSSM: 
\begin{eqnarray}
\dot s_{12}&=&-c_{12}(s_{12}s_{23}-c_{12}c_{23}s_{13})
(-c_{12}s_{23}-s_{12}c_{23}s_{13})\kappa A_{21}y^2_\tau\nonumber\\
&-&s_{12}c_{12}c_{23}s_{13}(s_{12}s_{23}-c_{12}c_{23}s_{13})\kappa A_{31}
y^2_\tau\label{eqn:runs12}\\
&+&c_{12}^2c_{23}s_{13}(-c_{12}s_{23}-s_{12}c_{23}s_{13})\kappa A_{32}y^2_\tau
\nonumber 
\end{eqnarray}
\begin{eqnarray}
\dot s_{23}&=&s_{12}c_{23}^2(s_{12}s_{23}-c_{12}c_{23}s_{13})\kappa A_{31}
y^2_\tau\nonumber\\
&-&c_{12}c_{23}^2(-c_{12}s_{23}-s_{12}c_{23}s_{13})\kappa A_{32}y^2_\tau
\label{eqn:runs23}
\end{eqnarray}
\begin{eqnarray}
\dot s_{13}&=&-c_{12}c_{23}c_{13}^2(s_{12}s_{23}-c_{12}c_{23}s_{13})
\kappa A_{31}y^2_\tau\nonumber\\
&-&s_{12}c_{23}c_{13}^2(-c_{12}s_{23}-s_{12}c_{23}s_{13})\kappa A_{32}
y^2_\tau. \label{eqn:runs13}
\end{eqnarray}

The evolution of the mixing angles can be classified into several universal
types of behaviour, depending on the magnitude of the factors $A_{ab}$ 
in eqs.~(\ref{eqn:runs12}-\ref{eqn:runs13}). We note that, neglecting the
small effects of mass evolution, all possible mass configurations in patterns 
$i)$-$iii)$ listed at the end of Section 2 (except for the case of neutrinos 
(approximately) degenerate in mass and all having the same CP parity - to be 
discussed later)  give one of the following four structures:
\begin{itemize}
\item[a)] $A_{31}\approx A_{32}$ and $|A_{31}|\approx |A_{21}|\approx1$
\item[b)] $A_{31}\approx A_{32}$ and $|A_{21}|\gg|A_{31}|$, $|A_{21}|\gg1$
\item[c)] $A_{32}\approx A_{21}\approx0$, $|A_{31}|\gg1$
\item[d)] $A_{31}\approx A_{21}\approx0$, $|A_{32}|\gg1$
\end{itemize}

For hierarchical masses and for partly degenerate structure with opposite CP 
parities of the (almost) degenerate neutrinos all $A_{ab}$ are ${\cal O}(1)$. 
For partly degenerate pattern with same CP parities or for degenerate pattern,
at most one of them is large. In the first case, it follows from 
eqs.~(\ref{eqn:runs12})-(\ref{eqn:runs13}) and the value of $\epsilon$
ranging from $10^{-5}$ in the SM to $2.5\times10^{-2}$ in the MSSM with
$\tan\beta\approx50$ that the evolution of the angles is very weak
\cite{ELLO,HAOK,CHKRPO,CAESIBNA4}. On the other hand, for one of the $A_{ab}$ 
factors sufficiently big so that $|A_{ab}\epsilon|\simgt1$, the angles evolve 
to an infrared quasi-fixed point (FP) \cite{CHKRPO,CAESIBNA4}. It is clear 
{}from eqs.~(\ref{eqn:runs12})-(\ref{eqn:runs13}) and from the parametrization 
(\ref{eqn:MNSmat}) of the MNS matrix $U$, that depending on which $A_{ab}$ is 
large, the fixed points are either at $U_{31}=0$ or $U_{32}=0$. Before 
discussing the approach to those fixed points in more detail, we can already 
now summarize several qualitative conclusions.

It is interesting to notice that in both fixed points we get the 
same relation between the mixing angles
 \begin{eqnarray}
\sin^22\theta_{12} =
{s^2_{13}\sin^22\theta_{23}\over(s_{23}^2c^2_{13} + s^2_{13})^2}
 ~~~~{\rm with} 
 ~~~~s_{23}^2 = {1\over2}\left(1\pm\sqrt{1-\sin^22\theta_{23}}\right). 
\label{eqn:FPrel}
\end{eqnarray}
Thus, contrary to the $2\times2$ case, quantum corrections can give now an 
interesting, non-trivial fixed point relation. It is particularly interesting 
in the context of the present experimental indications for small $\theta_{13}$ 
angle from CHOOZ and maximal atmospheric and solar mixing
($\sin^22\theta_{23}\approx1$ and $\sin^22\theta_{12}\approx1$). The relation 
(\ref{eqn:FPrel}) is inconsistent with such a pattern of mixing. We stress 
that quantum corrections summarized in RGEs 
eqs.~(\ref{eqn:runs12})-(\ref{eqn:runs13}), if large, always give 
(\ref{eqn:FPrel}). Thus, if the presently most likely pattern of mixing 
is confirmed experimentally, all mass patterns
generating large quantum corrections through RGE are ruled out!

Special cases easy to consider (still before a detailed study of the 
approach to the fixed points) are exact degeneracies at the scale $M$: 
$m_{\nu_1}^2=m_{\nu_2}^2\neq m_{\nu_3}^2$ or 
$m_{\nu_1}^2=m_{\nu_2}^2= m_{\nu_3}^2$. As follows from the discussion
surrounding eq.~(\ref{eqn:crosscond}), for the same CP parities, 
$m_{\nu_1}=m_{\nu_2}$ or $m_{\nu_1}=m_{\nu_3}$ or $m_{\nu_2}=m_{\nu_3}$ the 
angles must satisfy the FP relation (\ref{eqn:FPrel}) already at the scale 
$M$. If this is ruled out by experiment, then the two-fold degeneracy needs
$m_{\nu_1}=-m_{\nu_2}$ and the three-fold degeneracy is ruled out. This
conclusion holds (both in the SM and MSSM) under the assumption that the 
RGE corrections are the dominant ones (see next section for other 
possibilities). For $m_{\nu_1}=-m_{\nu_2}$ and $m^2_{\nu_1}\gg m^2_{\nu_3}$, 
with $|m^2_{\nu_3}-m^2_{\nu_2}|\approx\Delta M^2$, we can ask if quantum 
corrections can explain $\Delta m^2=|m^2_{\nu_2}-m^2_{\nu_1}|$. The answer 
to this question is positive in the MSSM \cite{BAROST} (in the SM see 
\cite{CAESIBNA3}) and will be discussed in more detail in Section 5.
Another point worth a discussion is what if experiment will eventually be 
consistent with the FP relation (\ref{eqn:FPrel}). Two-fold degeneracy 
is easy: we need $m_{\nu_1}=m_{\nu_2}$,
$|m^2_{\nu_3}-m^2_{\nu_2}|\approx\Delta M^2$ and (as discussed in Section 5)
quantum corrections can explain $\Delta m^2$. 
With three-fold degeneracy, we necessarily have at least one pair of neutrinos 
with the same CP parities but the question is whether we can explain both 
$\Delta m^2$ and $\Delta M^2$ by the discussed here class of quantum 
corrections. The answer is negative \cite{DIJO,CAESIBNA1,CAESIBNA2}. This 
can be explained as follows: from eq.~(\ref{eqn:solmass}) we would need 
$|U_{31}|\approx |U_{32}|$ (to keep $m^2_{\nu_1}\approx m^2_{\nu_2}$) during 
the entire evolution and, at the same time, $|U_{33}|-|U_{32}|\sim{\cal O}(1)$ 
to generate sufficiently large $\Delta M^2$. With $\theta_{13}\approx0$ the 
first condition implies $\sin^22\theta_{12}\approx1$ and the second 
$\sin^22\theta_{23}\approx1$ ($|U_{33}|-|U_{32}|$ can be at most $\sim1/2$).
This is however incompatible with the  relation (\ref{eqn:FPrel}) which should 
be satisfied! So, even if experiment was consistent with (\ref{eqn:FPrel}),
the three-fold degenerate mass spectrum would still be unacceptable since
$\Delta M^2$ and $\Delta m^2$ could not be explained by quantum corrections 
(always under the assumption about the dominance of the RG corrections).

Let us now discuss the approach to the fixed point and concentrate first on 
the structure b) which can be realized for partly degenerate or  degenerate 
patterns,
with same CP parities of $\nu_1$ and $\nu_2$. Since $A_{31}\approx A_{32}$ and
$|A_{21}|\gg|A_{31}|$, equations~(\ref{eqn:runs12}-\ref{eqn:runs13}) reduce to:
\begin{eqnarray}
\dot s_{12}&=& -c_{12}(s_{12}s_{23}-c_{12}c_{23}s_{13})
(-c_{12}s_{23}-s_{12}c_{23}s_{13})A_{21}y^2_\tau   -
c_{12}s_{23}c_{23}s_{13}A_{32}y^2_\tau,\nonumber\\                    
\dot s_{23}&=&s_{23}c_{23}^2A_{32}y^2_\tau,\label{eqn:runssred}\\   
\dot s_{13}&=&c^2_{23}s_{13}c_{13}^2A_{32}y^2_\tau\nonumber 
\end{eqnarray}
(we take MSSM, with $\kappa=1$). The equation for $\dot s_{23}$ is the 
same as in the $2\times2$ scenario discussed in the previous subsection. The 
evolution of $\theta_{23}$ in the two cases is formally not identical as the 
evolution of the mass factor $A_{32}$ depends now also on the remaining 
mixing angles $\theta_{12}$ and $\theta_{13}$. Nevertheless, the qualitative 
behaviour of $s_{23}$ is similar because in most cases the scale dependence 
of $A_{32}$ can be neglected. Denoting 
\begin{eqnarray}
\xi_\tau\equiv\exp\left(-\int_0^t
2A_{32}(t^\prime)y^2_\tau(t^\prime)dt^\prime\right)
\approx\exp\left(-2A_{32}(0)\epsilon\right) 
\end{eqnarray}
the solution for $s^2_{23}(t)$ reads 
\begin{eqnarray}
s^2_{23}(t) = s^2_{23}(0)/\left[s^2_{23}(0) + c^2_{23}(0)\xi_\tau\right] 
\label{eqn:sinatmexact}
\end{eqnarray}
and yields 
\begin{eqnarray}
\sin^22\theta_{23}(t)=\xi_\tau\sin^22\theta_{23}(0)/\left[s^2_{23}(0)  +
c^2_{23}(0)\xi_\tau\right]^2  . 
\label{eqn:sinatmeps}
\end{eqnarray}
The solution for $s^2_{13}$ can also be given in a closed form:
\begin{eqnarray}
s^2_{13}(t)=s^2_{13}(0)/\left\{s^2_{13}(0)+c^2_{13}(0)
\left[s^2_{23}(0)+c^2_{23}(0)\xi_\tau\right]\right\}.
\label{eqn:s13sol}
\end{eqnarray}
Thus, since $|A_{32}|\approx1$, the evolution 
of both, $s_{23}$ and $s_{13}$ is very weak \cite{HAOK,CHKRPO,CAESIBNA4}. 
For example, the effect of the running for $\sin^22\theta_{\rm atm}$ is a 
2.5\% change for extreme value of $\tan\beta\approx50$. 
 
For $|A_{21}\epsilon|\gg1$ (recall we consider $m^2_{\nu_{2(1)}}>|\Delta m^2|$ 
and $m_{\nu_1}m_{\nu_2}>0$), the evolution is towards one of the two 
approximate fixed points of the RG equation for $s_{12}$. One can easily 
check, for instance, by considering the equation for 
$d\tan\theta_{12}/dt$, that for $A_{21}>0$ (i.e. for
$\Delta m^2>0$) the point $U_{31}=0$ is the UV fixed point and $U_{32}=0$
is the IR fixed point. For $A_{21}<0$ (i.e. for $\Delta m^2<0$) the situation
is reversed. It is also interesting to notice that in the limit $s_{13}=0$
we can follow analytically the approach to  the fixed points. In this 
approximation
\begin{eqnarray}
\dot s_{12}=s_{12}c_{12}^2s_{23}^2A_{21}y^2_\tau
\label{eqn:s1fors3zero}
\end{eqnarray}
and the solution for $s^2_{12}$ is of the form~(\ref{eqn:sinatmexact}),
with $s_{23}(c_{23})\rightarrow s_{12}(c_{12})$ and\footnote{Eqs.
(\ref{eqn:s1fors3zero}) and (\ref{eqn:xiprime}), after solving for $s_{23}(t)$
and in the approximation of constant $A_{21}$ and $A_{23}$, give:
\begin{eqnarray}
s^2_{12}(t)=s^2_{12}(0)/\left\{s^2_{12}(0)+c^2_{12}(0)\left[
c^2_{23}(0)+s^2_{23}(0)\xi_\tau^{-1}\right]^{-A_{21}/A_{32}}\right\}.
\nonumber
\end{eqnarray}
} 
\begin{eqnarray}
\xi_\tau\rightarrow\xi_\tau^\prime=\exp\left(-\int_0^t
2s^2_{23}(t^\prime)A_{21}(t^\prime)y^2_\tau(t^\prime)dt^\prime\right). 
\label{eqn:xiprime}
\end{eqnarray}
For $A_{21}>0$, in the top-down running, the factor $\xi^\prime\rightarrow0$ 
exponentially with decreasing the scale $Q\rightarrow M_Z$ i.e. for growing
$t\propto A_{21}y^2_\tau\log(M/Q)$ and, consequently, we obtain 
$s_{12}(t)=\pm1$ (depending on its initial sign) and approach IR fixed point 
at $U_{32}=0$. Changing the signs of the right hand sides of 
eqs. (\ref{eqn:runssred}) one can see that in the bottom-up evolution
we approach $s^2_{12}(t)\approx0$ 
exponentially, i.e. the UV fixed point at $U_{31}=0$. For $A_{21}<0$ we get 
the reversed situation, in accord with our general expectations.

It is interesting to estimate the values of $A_{21}$ and $\tan\beta$,
for which the approach to the IR fixed points is seen. In ref.~\cite{CHKRPO}
we estimated that for approaching the fixed point during the evolution in the 
range $(M,M_Z)$ with $M\approx10^{10}$~GeV one needs $A_{21}\epsilon(M)>3$, 
i.e. for $\tan\beta=20$ one needs $m_{\nu_1}\approx m_{\nu_2}\simgt10^{-4}$~eV 
for $\Delta m^2\sim10^{-10}$ eV$^2$ and 
$m_{\nu_1}\approx m_{\nu_2}\simgt0.01$ eV for 
$\Delta m^2\simgt10^{-6}$ eV$^2$. The qualitative change from a negligible
evolution to the FP behaviour at the electroweak scale is abrupt and occurs 
in the small range $0.5\simlt|A_{ab}\epsilon|\simlt3$.

Finally we note that, from the point of view of the initial conditions
at the scale $M$, the UV fixed point looks not realistic as the neglected
muon Yukawa coupling $y_\mu$ quickly destabilizes it during the evolution.
We conclude that for hierarchical and partly degenerate mass patterns
the evolution of the mixing angles is either very mild or shows (for
$|A_{21}|\epsilon\simgt3$) a fixed point behaviour. 

The evolution of mixing angles in the degenerate case, 
$m_{\nu_3}^2\approx m_{\nu_2}^2\approx m_{\nu_1}^2\sim{\cal O}(\Delta M^2)$ 
or larger, partly falls  into the same classes of behaviour. Indeed, as 
long as $A_{31}\approx A_{32}$ with  $|A_{31}|\simlt{\cal O}(1)$
and $|A_{21}|\gg1$, the 
angles evolve according to the same equations (\ref{eqn:runssred}). One can 
easily identify the mass patterns of the degenerate case that fall into 
this category: the necessary condition is that $m_{\nu_1}$ and $m_{\nu_2}$ 
are of the same sign. For the evolutions of $s_{12}$ we then closely follow 
the two possibilities, depending on the sign of $A_{21}$, discussed for the 
partial degeneracy with $m_{\nu_1}m_{\nu_2}>0$. We simply note that larger 
values of $|A_{21}|$ are generic for the present case and the approach to the 
fixed points is faster. However, the evolution of $s_{23}$ and $s_{13}$ are 
guaranteed to be mild only if $m_{\nu_1}$ and $m_{\nu_2}$ are negative. For 
positive $m_{\nu_1}$ and $m_{\nu_2}$, i.e. when all masses have the same CP 
parity, we can have $|A_{32}\epsilon|\simgt3$ for $\tan\beta>40$
(due to the bound (\ref{eqn:2bb0nubound}) 
$|A_{32}\epsilon|\simgt3$ cannot be realized for $\tan\beta<40$). 
The angle $\theta_{23}$ behaves then as the angle $\vartheta$ of the 
$2\times2$ mixing discussed in Section 4.3 and, according to the solution 
(\ref{eqn:sinatmexact}),
$s_{23}$ is exponentially focused to the stable FP $s_{23}(t)=0$
or $s_{23}(t)=\pm1$, depending on the sign of $A_{32}$, and on the direction
of the evolution. The angle $s_{13}$ behaves in a similar way except
that, as follows from eq.~(\ref{eqn:s13sol}), it does not reach the value
$s^2_{13}=1$ when $s^2_{23}\rightarrow1$ (due to the presence of the factor
$c^2_{23}$ in its RGE, the evolution of $s_{13}$ is then ``frozen''). We 
conclude that in the regime in which the approach to the fixed points is 
relevant, the pattern with approximately degenerate neutrino masses and
the same all three CP parities is not acceptable.

As we said earlier, the approach to the FP behaviour is abrupt as a function
of $|A_{31}\epsilon|\approx|A_{32}\epsilon|$. However, in the small transition
region of the values of $A_{31}$ and $A_{32}$, in agreement with our 
discussion in the previous subsection, it is possible to chose the initial 
condition for $s_{23}$ so to get $\sin^22\theta_{23}=1$, $s_{13}\approx0$ 
at the electroweak scale. This was exploited in ref.~\cite{BADIMOPA2} as a 
possible mean to obtain maximal atmospheric neutrino mixing from the initially 
small $\sin^22\theta_{23}$. Since $|A_{21}\epsilon|$ is always much larger 
than $|A_{32}\epsilon|$, the evolution of $s_{12}$ is then such that the FP 
at $U_{32}=0$ or $U_{31}=0$ is quickly reached. Thus, a realistic solution, 
with maximal $\theta_{23}$ and $s_{13}\approx0$, has 
$\sin^22\theta_{12}\approx0$ and, moreover, the scheme cannot work unless 
there is an extreme fine tuning of the initial parameters \cite{CAESIBNA4}.

The remaining degenerate mass patterns can be classified according to the
relations $A_{32}\approx A_{21}\approx0$ or $A_{31}\approx A_{21}\approx0$.
Consider first $A_{21}\approx A_{32}\approx0$, i.e. 
$m_{\nu_1}\approx-m_{\nu_2}\approx m_{\nu_3}$. The 
equations for the evolution of the mixing angles can be approximated as
\begin{eqnarray}
\dot s_{12} &=& -s_{12}c_{12}c_{23}s_{13}
(s_{12}s_{23}-c_{12}c_{23}s_{13})A_{31}y^2_\tau,\nonumber\\
\dot s_{23} &=& s_{12}c^2_2
(s_{12}s_{23}-c_{12}c_{23}s_{13})A_{31}y^2_\tau,\label{eqn:reda}\\
\dot s_{13} &=& -c_{12}c_{23}c^2_3
(s_{12}s_{23}-c_{12}c_{23}s_{13})A_{31}y^2_\tau.\nonumber
\end{eqnarray}
These equations exhibit IR quasi-fixed point behaviour for 
$A_{31}\epsilon\ll-1$, corresponding to $U_{31}=0$. As before, at the fixed 
point the angles satisfy the relation $s_{13}=\tan\theta_{12}\tan\theta_{23}$.
Since $\dot s_{12}$ is proportional to $s_{12}$ and suppressed by $s_{13}$, 
the running of $s_{12}$ is weak. The IR fixed point is reached due to strong 
running of $s_{23}$ and $s_{13}$. For $A_{31}\epsilon\gg1$, $s_{23}(M_Z)$ 
is strongly focused at $\pm1$. Thus, the mass and $\tan\beta$ configurations 
leading to $A_{31}\epsilon\gg1$ are unacceptable. For $A_{31}\approx 
A_{21}\approx0$ and $A_{32}<0$ we get IR fixed point in $U_{32}=0$.

The RG running of the mixing of three neutrinos for a complex matrix $U$, 
i.e. non-zero phase $\delta$ and $\alpha_1$, $\alpha_2\neq0$ modulo $\pi/2$ 
with all masses $m_{\nu_a}$ positive by definition, has been also investigated 
in the literature \cite{CAESIBNA4}. For initially degenerate $\nu_a$ and $\nu_b$
the mixing pattern is determined by the condition 
${\rm Re}\left(U^{3a}U^{3b\ast}\right)\equiv 
{\rm Re}\left(U^{3a}_{\rm MNS}U^{3b\ast}_{\rm MNS}\right)=0$ playing in 
the general complex case the role of the FP condition. Qualitatively, our 
conclusions remain unchanged also in this case. For example, starting with
$m_{\nu_1}^2=m_{\nu_1}^2$, $\sin^22\theta_{23}=1$, $s_{13}\approx0$ and 
arbitrary phases $\alpha_1$, $\alpha_2$ at the scale $M$, the RG evolution 
leads to the relation \cite{CAESIBNA4}
\begin{eqnarray}
\sin^22\theta_{12}(M_Z)=\sin^22\theta_{12}(M)\sin^2(\alpha_1-\alpha_2)+
{\cal O}(s^2_{13})\label{eqn:FPcomplex}
\end{eqnarray}
We see therefore that maximal solar mixing can again be obtained only if
the change of $\theta_{12}$ during the running is negiligible, i.e. for
$\alpha_1-\alpha_2\approx\pi/2$ at the scale $M$.

The running of the masses is always given
by eq.~(\ref{eqn:massrge}) and is similar as for a real matrix $U$.
The conclusion that the triple degeneracy of neutrinos at the scale $M$,
$|m_{\nu_1}|=|m_{\nu_2}|=|m_{\nu_3}|$, cannot lead to an acceptable pattern 
of masses and mixing also remains valid. 

In summary, with all $|A_{ab}\epsilon|\simlt0.5$ the evolution of the 
mixing is negligible. For the mass configurations such that at least one 
$|A_{ab}|\gg1$ and $|A_{ab}\epsilon|\simgt3$ the infrared fixed points are 
reached during the evolution, independently of further details of the mass 
matrices. However, only for $|A_{21}\epsilon|\simgt3$ and 
$|A_{31}|, |A_{32}|\simlt1$, or for $A_{31}(A_{32})\epsilon\ll-3$ and 
$A_{31}(A_{32})\approx A_{21}\approx0$ the 
evolution is consistent with a large atmospheric mixing angle at low energy. 
The mass configurations $m_{\nu_1}\approx m_{\nu_2}\approx m_{\nu_3}$ and 
$\pm m_{\nu_1}\approx\mp m_{\nu_2}\approx m_{\nu_3}$ with $A_{31}(A_{32})>0$
also lead to the infrared fixed points but at the same time the atmospheric
mixing angle converges to zero. We also note that for 
$|A_{21}\epsilon|\simgt3$ only 
$s_{12}$ runs to assure the fixed point relation, so the initial values for 
$s_{23}$ and $s_{13}$ have to be close to their experimental values already 
at the scale $M$. For $A_{31}\epsilon\simlt-3$ or $A_{32}\epsilon\simlt-3$, 
$s_{23}$ and $s_{13}$ 
evolve strongly and the evolution of $s_{12}$ is weak. The IR fixed point 
relation (\ref{eqn:FPrel}) is always one equation for three angles. Insisting 
on a large atmospheric mixing angle, it correlates small (as follows from 
CHOOZ) $\theta_{13}$ angle with a small solar mixing angle. The fixed point 
solution makes the low energy angles dependent on only two, instead of in 
general three, initial conditions at the scale $M$. We conclude that quantum 
corrections encoded in the RG running of mixing angles may have dramatic 
impact on their physical values if the mass pattern is partial degeneracy
or degeneracy. If the bimaximal mixing solution was confirmed, all the mass
patterns leading to the FP would be ruled out, 
unless the low energy threshold corrections change the results.

\section{Low energy threshold corrections}
\setcounter{equation}{0} 

\subsection{Threshold corrections in the SM}

The Wilson coefficient $\mbox{\boldmath$C$}^{AB}$ of the dimension 5
$\Delta L=2$ operator (\ref{eqn:dim5op}) is a renormalized parameter of
the effective theory Lagrangian. Integrating its RGEs from the high scale
$M$ down to some scale $Q\approx M_Z$ allows to resum potentially large
corrections involving $\ln(M/Q)$ to all orders of the perturbation 
expansion. However, since the low energy scale $Q$ is not a priori 
determined by any physical requirement (apart from the condition 
$Q\approx M_Z$), the neutrino masses and mixing angles computed in 
the tree-level approximation from the Wilson coefficient 
$\mbox{\boldmath$C$}^{AB}(Q)$ do depend (albeit weakly) on the
actual choice of $Q$. This dependence can be removed by computing  
masses and mixing angles in the one-loop approximation in the MS scheme
and with $Q$ as the renormalization scale. 

In general, to compute the physical neutrino masses and mixing one has 
first to perform the RG evolution of the entire matrix $U$ and neutrino 
masses from the scale $M$ down to the scale $M_Z$, include threshold 
corrections and subsequently rediagonalize the resulting mass matrix 
$m_{ab}^{\rm 1-loop}$ by an additional unitary matrix $U^\prime$. The 
physical matrix $U_{\rm MNS}$ (whose elements are probed in neutrino 
experiments) is then given as 
$e^{i\varphi_A}U_{\rm MNS}^{Aa}=(U\cdot U^\prime)^{Aa}$ where $U$ is the 
matrix obtained from the RG evolution. 

In the basis in which neutrino masses are diagonal, we have
\begin{eqnarray}
m_{ab}^{\rm 1-loop}=m_{\nu_a}\delta^{ab}+m_{\nu_a}I^{ab}+m_{\nu_b}I^{ba}
\label{eqn:moneloop}
\end{eqnarray}
where 
\begin{eqnarray}
I^{ab}=\sum_{A,B}U^{Aa\ast}I_{AB}^{\rm th}U^{Bb}
\label{eqn:Idef}
\end{eqnarray}

In the SM one finds \cite{CHWO}
\begin{eqnarray}
I_{AB}^{\rm th}={\delta^{AB}\over16\pi^2}{g^2_2\over2}
{m_{e_A}^2\over M^2_W}\left[{11\over8}-{3\over2}\ln{M_W\over Q} 
+{\cal O}\left(x_A\ln x_A\right)\right]+\dots
\label{eqn:SMthcor}
\end{eqnarray}
where $x_A\equiv m^2_{e_A}/M^2_W$. Note that since 
$g^2_2m_{e_A}^2/2M^2_W=y^2_{e_A}$, the coefficient of $\ln(1/Q)$ 
agrees with the coefficient $\kappa$ of the $y^2_{e_A}\mbox{\boldmath$C$}^{AB}
+\mbox{\boldmath$C$}^{AB}y^2_{e_B}$ term in the RGE (\ref{eqn:rgeCgen}) for
the SM. This confirms the correctness of the recent re-derivation 
\cite{ANDRKELIRA} of the SM RGE. The remaining corrections indicated by dots 
in (\ref{eqn:SMthcor}) affect only the overall scale of the neutrino
masses and therefore are not interesting in view of the unspecified magnitude
of the mass $M$ in eq.~(\ref{eqn:diagmass}). 

As could be expected, in the SM the nontrivial part (\ref{eqn:SMthcor}) of 
the low energy threshold corrections that changes the structure of the mass 
matrix, and hence of the matrix $U^{Aa}$, is small and always negligible 
compared to the quantum effects described by the RG running, which are enhanced
by large logarithm of the ratio $M/M_Z$. The correction (\ref{eqn:SMthcor}) 
can be most easily taken into account by stopping the RG evolution of the 
Wilson coefficient $\mbox{\boldmath$C$}^{AB}$ at the scale $Q=M_We^{-11/12}$.

\subsection{Threshold corrections in the MSSM}

In the supersymmetric model, apart from stabilizing the results obtained from
the RG analysis, the low energy threshold corrections can be as or more 
important than the RG evolution and may have very important physical 
consequences for the neutrino masses and mixing angles 
\cite{CHUPO,CHU,CHIOPOVA}. Before we discuss some physical examples, it is 
worthwhile to adapt our calculational procedure to the new situation, so that
the RG evolution effects and the threshold corrections can be treated on
equal footing. Although the general procedure outlined earlier is in principle
correct irrespectively of the relative magnitude of the RG and threshold 
effects, in practice it can mask simple qualitative features if the 
rediagonalization due to threshold corrections is not a small perturbation.

We observe that, to a very good accuracy, in the solution (\ref{eqn:ellosol}) 
to the RGE the factors $I_{e_A}$ (\ref{eqn:lolaIfact}) can be approximated by
\begin{eqnarray}
I_{e_A}\approx1-\kappa\int_0^{t_Z}y^2_{e_A}dt^\prime\equiv1-I^{\rm rg}_A.
\end{eqnarray}
With this approximation (which can fail only for unrealistically large 
values of $\tan\beta$)
all quantum effects of the physics below the scale $M$, the RG running
as well as the low energy threshold corrections, can be described by a single
formula (\ref{eqn:moneloop}). The $m_{\nu_a}$ and the matrix $U$ are now
the neutrino mass eigenvalues and the neutrino mixing matrix, respectively, 
at the scale $M$, and the factors $I^{ab}$ are given by
eq.~(\ref{eqn:Idef})with
\begin{eqnarray}
I_{AB}^{\rm th}\rightarrow I_{AB}^{\rm th}
-\delta^{AB}I^{\rm rg}_A\equiv I_{AB}.
\end{eqnarray}

Again, the case of a real matrix $U$ is particularly easy because then 
the formula (\ref{eqn:moneloop}) can be written as
\begin{eqnarray}
m_{ab}^{\rm 1-loop}=m_{\nu_a}\delta^{ab}+(m_{\nu_a}+m_{\nu_b})I^{ab}
\equiv m_{\nu_a}\delta^{ab}+\Delta m_{ab}
\label{eqn:moneloop2}
\end{eqnarray}
with the right hand side symmetric and real i.e. Hermitean. One can then use 
the formal perturbation calculus of quantum mechanics (see e.g. 
\cite{SCHIFF}) to find corrections to neutrino masses and mixing angles. 
Of course, if the threshold corrections are absent\footnote{If the threshold 
corrections are universal, i.e. if $I_{AB}^{\rm th}=I^{\rm th}\delta^{AB}$, 
they can be absorbed into the overall scale of neutrino masses and do not 
influence neither the mixing angles nor the ratio $\Delta M^2/\Delta m^2$.}
one has $I_{AB}\approx\delta^{AB}I_A$ with $|I_\tau|\gg|I_\mu|\gg|I_e|$.

Several physically interesting situations can be discussed. In the following 
we shall mainly focus on the scenarios with equal three or at least two masses
at the scale $M$. Although the discussion of the mixing angles does not
depend on whether the masses are equal or only approximately equal, the former
case is much more constrained for the masses themselves and therefore more
interesting. 

\vskip0.5cm
\noindent {\bf 5.2.1 Three-fold degeneracy and flavour diagonal corrections}
\vskip0.3cm

\noindent In the MSSM the complete formulae for $I_{AB}^{\rm th}$ are lengthy 
due to additional contributions of $H^\pm$ and chargino/charged slepton and 
neutralino/sneutrino loops \cite{CHWO} but we do not need them for our 
discussion. First of all, it is interesting to investigate whether it is 
possible to generate the right mass squared differences starting from 
degenerate neutrinos at the scale $M$: $|m_{\nu_1}|=|m_{\nu_2}|=|m_{\nu_3}|$ 
(we again allow for different CP parities of the neutrinos). From the 
arguments presented in Section 4 it follows that this is impossible if the 
factors $I_{AB}$ are dominated by the RG effects, i.e. when 
$I_{AB}\approx\delta^{AB}I_A$ with $|I_\tau|\gg|I_\mu|, |I_e|$. The degeneracy 
Ansatz is however potentially interesting as it might help to understand the 
bimaximal mixing. Also, the neutrino masses of order of few eV are
necessary if the neutrinos are to play a role of the hot component of the 
Dark Matter.\footnote{The hot component of the Dark Matter was previously 
needed to account for formation of largest scale structures in the Universe. 
At present, in view of the observational evidence for a significant 
contributions of the cosmological constant (or another form of dark energy 
as e.g. quintessence) to the energy density of the Universe, the hot component 
of the Dark Matter seems no longer necessary (but is not excluded).}
It is therefore interesting to see if some
other corrections can split the mass squares.

We start with the flavour diagonal threshold corrections: 
$I_{AB}=\delta^{AB}I_A$ and consider first the mass pattern 
$m_{\nu_1}=-m_{\nu_2}=m_{\nu_3}\equiv m_{\nu}$.
The formula (\ref{eqn:moneloop2}) now takes the form
\begin{eqnarray}
m_{ab}^{\rm 1-loop}=m_{\nu}\left(
\matrix{1+2U^2_{A1}I_A &       0        & 2U_{A1}U_{A3}I_A\cr
             0         &-1-2U^2_{A2}I_A &        0         \cr
       2U_{A1}U_{A3}I_A&       0        & 1+2U^2_{A3}I_A}\right)
\end{eqnarray}
The corrections to neutrino masses can be calculated perturbatively. Because 
$m_{\nu_1}=m_{\nu_3}$, one applies here the perturbation calculus to the case 
with degeneracy of the unperturbed ``energy levels'' \cite{SCHIFF}. Instead of
solving the ``secular'' equation for the corrected eigenvalues, it is however 
better to exploit the freedom of an arbitrary rotation in the plane $(13)$: 
$U\rightarrow UR_{13}$. This freedom can be used to diagonalize the 
perturbation by requiring
\begin{eqnarray}
\sum_AU_{A1}U_{A3}I_A=0\label{eqn:diagcond}
\end{eqnarray}
which fixes the matrix $U$ after taking into account the perturbation. 
The correction $\Delta m_{ab}$ to the zeroth order neutrino mass matrix 
becomes then diagonal and the neutrino masses are given by 
\begin{eqnarray}
|m_{\nu_a}| = m_{\nu}\left(1+2U^2_{Aa}I_A\right).\label{eqn:masseq}
\end{eqnarray}

Let us now assume \cite{CHIOPOVA} that 
$I_e\neq0$, $I_\mu=I_\tau=0$. The condition (\ref{eqn:diagcond}) 
reduces then to $U_{11}U_{13}=0$. The experimentally viable solution is 
$U_{13}=0$, i.e. $s_{13}=0$. Note that this relation replaces the FP relation
(\ref{eqn:FPrel}) of Section 4. For the neutrino masses one finds
\begin{eqnarray}
\Delta m^2_{21} = -4m^2_{\nu}\cos^22\theta_{12}I_e, ~~~
\Delta m^2_{32} = -4m^2_{\nu}s^2_{12}I_e.
\label{eqn:viablesol}
\end{eqnarray}
It is therefore possible to generate $\Delta M^2\gg\Delta m^2$ provided the 
solar mixing is nearly maximal. The same result is obtained starting from 
$-m_{\nu_1}=m_{\nu_2}=m_{\nu_3}\equiv m_{\nu}$. Thus, if the 
underlying theory gives one of these two degeneracies together with the 
bimaximal mixing pattern, the low energy threshold corrections can be 
responsible for assuring $s_{13}=0$ and generating the correct mass squared 
differences. 

Of course it is unrealistic to expect only one correction $I_A$ to be
nonzero. Note however, that the results for the mixing angles and the ratio 
$\Delta M^2/\Delta m^2$ are not altered by shifting all corrections $I_A$ 
by an overall additive constant: $I_A\rightarrow \tilde I_A\equiv I_A-I$.
Indeed, using unitarity of the matrix $U$, the off-diagonal entries in 
the rhs of eq.~(\ref{eqn:moneloop2}) can be always written as 
$\sum_AU_{Aa}U_{Ab}\tilde I_A$ ($a\neq b$) with arbitrary $I$. For the 
diagonal entries instead one has $1+2\sum_AU^2_{Aa}I_A=
1+2I+2\sum_AU^2_{Aa}\tilde I_A\approx(1+2I)(1+2\sum_AU^2_{Aa}\tilde I_A)$.
Hence, up to higher order terms, eq.~(\ref{eqn:moneloop2}) can 
be rewritten  with $\tilde I^{ab}=\sum_AU_{Aa}U_{Ab}\tilde I_A$ and $m_\nu$ 
replaced by $\tilde m_\nu\equiv(1+2I)m_\nu$. Thus, 
$I_\mu=I_\tau\neq0$ is equivalent to $I_\mu=I_\tau=0$.

It follows that $|I_e|\gg|I_\tau|$, $|I_\mu|$, equivalent to 
$|\tilde I_e|\gg|\tilde I_\tau|\neq0$, $\tilde I_\mu=0$, should give only 
small correction to the result (\ref{eqn:viablesol}) and $s_{13}=0$. For 
$m_{\nu_1}=-m_{\nu_2}=m_{\nu_3}$,
solving eq.~(\ref{eqn:diagcond}) (with $I_e=\tilde I_e$, $I_\mu=0$ and
$I_\tau=\tilde I_\tau$) one finds \cite{CHIOPOVA} that 
\begin{eqnarray}
&&s_{13}=-{s_{12}\over c_{12}}s_{23}c_{23}r +{\cal O}(r^2)\nonumber\\
&&\Delta m^2_{32}\approx-4\tilde m_\nu^2\tilde I_es^2_{12}\label{eqn:valle}\\
&&\Delta m^2_{21}\approx-4\tilde m_\nu^2\tilde I_e
\left[\cos2\theta_{12}(1-s^2_{23}r)+(1+2c^2_{12})s^2_{13}\right]
\phantom{aaa}\nonumber
\end{eqnarray}
where $r\equiv\tilde I_\tau/\tilde I_e$. Thus, obtaining small but non-zero
angle $\theta_{13}$ is also possible. To obtain the experimentally favoured
value of $\Delta M^2$ one must have 
$|\tilde m_\nu^2\tilde I_e|\approx1.6\times10^{-3}$~eV$^2$ i.e. 
$|I_e-I_\mu|\sim10^{-3}$ and $I_\mu\approx I_\tau$
for $m_\nu$ in the eV range. In the MSSM the required 
hierarchy $|I_e|\gg|I_\tau|$, $I_\tau\approx I_\mu$ can arise from the 
low-energy threshold corrections only as a result of non-universality 
of the left-handed charged slepton masses (and sneutrino masses).\footnote{In 
the $\tilde W^\pm$ loop approximation to $I^{\rm th}_{AB}$ it was estimated 
in ref. \cite{CHIOPOVA} that in order to get $\tilde I_e\sim10^{-3}$ and 
positive  one needs $M_{\tilde e_L}\approx1.7M_{\tilde\mu_L}$ (and 
$M_{\tilde\mu_L}\approx M_{\tilde\tau_L}$). However, from the full expression 
for $I^{\rm th}_{AB}$ it follows \cite{CHWO} that $\tilde I_e\sim10^{-3}$ and 
positive can be achieved only for relatively light first chargino 
($m_{C_1}\simlt300$~GeV) and $M_{\tilde\mu_L}\approx(1.2-1.6)M_{\tilde e_L}$.
For $M_{\tilde e_L}\approx(1.2-1.6)M_{\tilde\mu_L}$, which can be
realized in the inverted hierarchy models \cite{DUPOSA}, one gets
$\tilde I_e\sim-10^{-3}$ i.e. positive 
$\Delta m^2_{21}$ for (almost) maximal solar mixing.}
Moreover, for exactly maximal solar mixing, obtaining $\Delta m^2$ appropriate 
for the LAMSW solution requires $s^2_{13}\approx10^{-(2-3)}$. This is 
consistent with the CHOOZ data and is realized for $r\simlt0.1$, i.e. 
$|I_\tau-I_\mu|\simlt10^{-4}$. For $\tan\beta\simlt3$ such a small difference 
between the corrections $I_\tau$ and $I_\mu$ can be due to the RG effects 
in $I^{\rm rg}_\tau$ and does not require any mass splitting between $\mu_L$ 
and $\tau_L$. For larger values of $\tan\beta$, for which 
$I^{\rm rg}_\tau-I^{\rm rg}_\mu>10^{-4}$, some conspiracy between 
$I^{\rm th}_\tau-I^{\rm th}_\mu$ due to 
$M_{\tilde\mu_L}< M_{\tilde\tau_L}$ and $I^{\rm rg}_\tau-I^{\rm rg}_\mu$
is required. The amount of the necessary fine tuning grows, of course, with 
$\tan\beta$. For the VO solution to the solar neutrino problem, with 
$\sin^22\theta_{12}=1$, a more severe fine tuning is always necessary:
$\Delta m^2\sim{\cal O}(10^{-10}$~eV$^2)$ requires
$|I_\tau-I_\mu|\simlt10^{-7}$ which is always smaller than 
$I^{\rm rg}_\tau-I^{\rm rg}_\mu\simgt10^{-5}$. Thus obtaining the correct
$\Delta m^2$ for this solution requires a cancellation at least to one part 
per hundred between the RG and threshold effects. For 
$\Delta m^2\sim{\cal O}(10^{-10}$~eV$^2)$ one gets $s^2_{13}\simlt10^{-8}$.
Similar results are obtained also for $-m_{\nu_1}=m_{\nu_2}=m_{\nu_3}$. 

It can be checked \cite{CHU} that, for the initially degenerate neutrino 
masses, the patterns $m_{\nu_1}=-m_{\nu_2}=m_{\nu_3}$ or
$-m_{\nu_1}=m_{\nu_2}=m_{\nu_3}$ and $|I_e|\gg|I_\tau|$, $I_\tau\approx I_\mu$
are the only ones that can produce the required mass squared splittings with 
flavour diagonal threshold corrections. The necessary condition is that
the underlying theory valid 
above the scale $M$ gives the bimaximal mixing in the basis in which 
eq.~(\ref{eqn:diagcond}) is satisfied \cite{CHIOPOVA,CHU}.

\vskip0.5cm
\noindent {\bf 5.2.2 Three-fold degeneracy and flavour non-diagonal 
corrections}
\vskip0.3cm

\noindent Qualitatively new possibilities for mixing angles and for splitting 
initially degenerate neutrinos originate from nonzero off-diagonal
elements of the corrections $I^{\rm th}_{AB}$ in eq.~(\ref{eqn:Idef})
\cite{CHUPO,CHU}. This is possible in the MSSM if the slepton mass matrices
are not diagonal in the flavour space in the basis in which the leptonic 
Yukawa couplings are diagonal. The amount of flavour mixing is then best 
quantified in terms of the so-called mass insertions \cite{GAGAMASI}
defined as the ratio of the off-diagonal (in flavour space) elements of the
charged sleptons mass squared matrices to some average charged slepton mass
squared.\footnote{Due to the underlying $SU_L(2)$ symmetry, mass insertions
inducing transitions between left-handed sleptons of different generations
are always accompanied by the insertions inducing similar transitions between
sneutrinos.} Current limits on such mass insertions 
following from the non-observation of the decays $\mu\rightarrow e\gamma$
etc. are not very stringent: Only the insertion $\delta_{LR}^{12}$
mixing left(right)-handed $\tilde\mu$ with right(left)-handed $\tilde e$ has 
to be smaller than ${\cal O}(10^{-5})$. The insertions $\delta_{LR}^{13}$ and
$\delta_{LR}^{23}$ causing similar, chirality changing,
$\tilde e\leftrightarrow\tilde\tau$ and  $\tilde\mu\leftrightarrow\tilde\tau$ 
transitions are bounded by $\approx0.5$ and $\approx0.1$, respectively,
for slepton masses $\sim500$ GeV. Bounds on the chirality preserving
insertions $\delta_{LL}^{AB}$, $\delta_{RR}^{AB}$
exist only for the $\tilde e\leftrightarrow\tilde\mu$ transition
and are $\approx0.2$ for $M_{\tilde l}\sim500$ GeV. The other
chirality preserving mass insertions are practically unrestricted.

If the mass insertions are non-vanishing, there is a flavour non-diagonal
contribution to $I^{\rm th}_{AB}$. In general it takes the form
\begin{eqnarray}
I^{\rm th}_{AB}\approx{1\over16\pi^2}\sum_{X,Y=L,R}
\delta_{XY}^{AB} h_{XY}(M_{\tilde l_{\rm av}},m_{C_j},m_{N_i})
\end{eqnarray}
where $h_{LL}$, $h_{RR}$ and $h_{LR}$ are some functions of the chargino 
and neutralino masses $m_{C_j}$, $m_{N_i}$ and of some average mass 
$M_{\tilde l_{\rm av}}$ of charged sleptons. The largest values of 
$I^{\rm th}_{AB}$ can be obtained for relatively light charginos, 
heavy sleptons ($\sim1$~TeV) and $M_2/\mu\approx-1$ and reach
$h_{LL}/16\pi^2\simlt{\rm few}\times10^{-3}$ ($h_{RR}$ and $h_{LR}$ are 
always smaller) \cite{CHWO}. For comparable chargino and slepton masses 
one has $h_{LL}/16\pi^2\simlt2\times10^{-4}$. In principle the mass 
insertion approximation should fail for $|\delta_{XY}^{AB}|\simlt0.1$. 
In practice it works as an order of magnitude estimate even for 
$|\delta_{XY}^{AB}|\simlt1$ (the error is then of order 25\%).

In the presence of non-zero mass insertions and 
for $m_{\nu_a}=m_{\nu_b}$ $(=-m_{\nu_c})$ the condition for vanishing of
the appropriate off-diagonal entry of the correction to the zeroth order 
mass matrix reads
\begin{eqnarray}
\sum_{A,B} U_{Aa}U_{Bb}I_{AB}=0.\label{eqn:cond}
\end{eqnarray}
Consider first the situation in which the single correction 
$I^{\rm th}_{e\mu}$,
$I^{\rm th}_{e\tau}$ or $I^{\rm th}_{\mu\tau}$ dominates over all other 
corrections. The condition (\ref{eqn:cond}) gives then relations 
between the mixing angles that are different from (\ref{eqn:FPrel}) and are 
listed in Table~\ref{tab:tab1} \cite{CHUPO}. Only three of the nine 
possibilities are compatible with the bimaximal mixing: dominant 
$I^{\rm th}_{\mu\tau}$ for $m_{\nu_1}=m_{\nu_3}$ or $m_{\nu_2}=m_{\nu_3}$ and 
dominant $I^{\rm th}_{e\tau}$ for $m_{\nu_1}=m_{\nu_2}$. For 
initially degenerate neutrinos, the latter combination gives wrong 
relation $\Delta m^2\approx2\Delta M^2$. (Other six combinations leading 
through (\ref{eqn:cond}) to $\sin^22\theta_{12}\sim\sin^22\theta_{13}$
also give bad relations $\Delta m^2\approx\Delta M^2$ or $\Delta M^2=0$.)
The former two are however interesting giving
$\Delta m^2\approx-4m^2_\nu\cos2\theta_{12}\sin2\theta_{23}
I^{\rm th}_{\mu\tau}$ and 
$\Delta M^2\approx4m^2_\nu(1+c^2_{12})\sin2\theta_{23}I^{\rm th}_{\mu\tau}$. 
Obtaining $\Delta M^2\approx3.2\times10^{-3}$~eV$^2$ is therefore
possible but only for $m_\nu\simgt1$~eV and $\delta^{23}_{LL}\simgt0.5$ i.e. 
for rather large flavour mixing in the slepton mass matrices.

\begin{table}[thb]
\caption[]{Relations of the FP type
between the mixing angles for dominant correction 
$I^{\rm th}_{AB}$ \label{tab:tab1}}
\vspace{0.2cm}
\begin{center}
\begin{tabular}{|c||c|c|c|}
\hline
\phantom{a} & $m_{\nu_1}=m_{\nu_2}$ & $m_{\nu_1}=m_{\nu_3}$ 
& $m_{\nu_2}=m_{\nu_3}$ \\
\hline
\hline
$I^{\rm th}_{e\mu}$ & $s_{13}=\cot2\theta_{12}\cot\theta_{23}$ 
            & ${s^2_{13}\over1-2s^2_{13}}=\cot\theta_{12}\tan\theta_{23}$ 
            & ${s^2_{13}\over1-2s^2_{13}}=-\tan\theta_{12}\tan\theta_{23}$ \\
\hline
$I^{\rm th}_{e\tau}$ & $s_{13}=-\cot2\theta_{12}\tan\theta_{23}$
            & ${s_{13}\over1-2s^2_{13}}=-\cot\theta_{12}\cot\theta_{23}$
            & ${s_{13}\over1-2s^2_{13}}=\tan\theta_{12}\cot\theta_{23}$ \\
\hline
$I^{\rm th}_{\mu\tau}$ 
            & ${2s_{13}\over1+s^2_{13}}=\tan2\theta_{12}\tan2\theta_{23}$
            & $s_{13}=-\tan\theta_{12}\cot2\theta_{23}$
            & $s_{13}=\cot\theta_{12}\cot2\theta_{23}$ \\
\hline
\end{tabular}
\end{center}
\end{table}

Similarly as for a non-zero $\tilde I_e$ correction, the solar mass 
squared difference  $\Delta m^2\ll\Delta M^2$ can be generated either by an 
appropriately tuned departure of the angle $|\theta_{12}|$ from maximal value 
$\pi/4$ or by another, hierarchically smaller, correction $I\neq0$. 
It has been demonstrated \cite{CHU}, that including 
on the top of the dominant $I^{\rm th}_{\mu\tau}$ correction a hierarchically 
smaller perturbation in the form of either $\tilde I_\mu$, $\tilde I_\tau$ 
($\tilde I_e$ does not work) or $I^{\rm th}_{e\mu}$, $I^{\rm th}_{e\tau}$ 
allows to split $m_{\nu_1}=-m_{\nu_2}$ even for exactly bimaximal 
mixing.\footnote{We note \cite{CHU}, that with a large $I_e$ diagonal 
perturbation discussed in the Subsection 5.2.1, 
$\Delta m^2\ll\Delta M^2$ (and $0\neq s^2_{13}\ll1$)
can be also induced by $I^{\rm th}_{e\mu}$ or $I^{\rm th}_{e\tau}$
instead of $I=\tilde I_\tau$ (small 
$I^{\rm th}_{\mu\tau}\neq0$ would not split $m_{\nu_1}$ and $m_{\nu_2}$
for $\sin^22\theta_{12}=1$). This leads to $s_{13}\approx-rs_{23}(c_{23})$
($r\equiv I^{\rm th}_{e\mu(\tau)}/\tilde I_e$) 
and $\Delta m^2\approx4m^2_\nu\sin2\theta_{12}c_{23}(s_{23})
I^{\rm th}_{e\mu(\tau)}$. Obtaining $\Delta m^2\sim{\cal O}(10^{-4}$~eV$^2$)
requires, for $m_\nu\sim1$~eV the correction 
$I^{\rm th}_{e\mu(\tau)}\sim10^{-(4-5)}$ which is possible in the MSSM
and gives $r\sim10^{-(1-2)}$ i.e. acceptable $s_{13}$.} 

In all those cases of two hierarchically different corrections
the important difference between the two alternatives: perturbation by a 
diagonal correction $\tilde I_A$ or perturbation by an off-diagonal correction
$I^{\rm th}_{AB}$ is that, for exactly bimaximal mixing, in the former case 
$\Delta m^2\propto r^2$, whereas in the latter $\Delta m^2\propto r$ only 
($s_{13}\sim r$ in all cases), where $r\ll1$ is the ratio of the smaller to 
larger correction as in (\ref{eqn:valle}) \cite{CHU}. Therefore, 
similarly as for $|\tilde I_e|\gg|\tilde I_\tau|$, obtaining 
$\Delta m^2$ appropriate for the VO solution with
$I^{\rm th}_{\mu\tau}$ dominance and hierarchically smaller 
$I_{e\tau}$ or $I_{e\mu}$ would require some tuning of slepton masses 
to cancel too large a contribution of $I^{\rm rg}_\tau$ to $I_\tau$.

Finally, we remind the reader that the relations listed in Table 1 remain
approximately valid when the equalities of the masses are relaxed and replaced
by the corresponding approximate degeneracies. The listed relations play
then the role of the FP relations discussed in Section 4. They are satisfied 
at the electroweak scale irrespectively of the initial values of the angles.
The observed $\Delta m^2$ and $\Delta M^2$ can be obtained by adjusting the
initial values of only approximately equal masses. The role of the threshold
corrections $I^{\rm th}_{AB}$ is then the same as the role of the RG 
corrections. If large enough, they give one of the ``fixed points'' relations
of Table 1.

\vskip0.5cm
\noindent {\bf 5.2.3 Two-fold degeneracy and threshold corrections}
\vskip0.3cm

\noindent We 
can also discuss the effect of threshold corrections in the case of the 
two-fold degeneracy $m^2_{\nu_1}=m^2_{\nu_2}\equiv m^2_\nu\gg m^2_{\nu_3}$ 
or $m^2_{\nu_1}=m^2_{\nu_2}\ll m^2_{\nu_3}$. For 
$m_{\nu_1}=-m_{\nu_2}$ the $\Delta m_{12}$ off-diagonal entry in 
eq.~(\ref{eqn:moneloop2}) automatically vanishes
but the correction matrix $\Delta m_{ab}$ as the whole needs not be diagonal.
In the first order of the perturbation calculus 
\cite{SCHIFF}, the neutrino masses are then given by 
eq.~(\ref{eqn:masseq}) and receive also further corrections of order 
${\cal O}\left((\Delta m_{ab})^2/{\rm max}(|m_{\nu_3}|, |m_\nu|)\right)$.
The mixing angles also receive corrections of  
order ${\cal O}\left(\Delta m_{ab}/{\rm max}(|m_{\nu_3}|,|m_\nu|)\right)$, 
i.e. small if the hierarchy of neutrino masses is large. If only flavour 
diagonal threshold corrections are present the solar mass squared splitting 
is given by\footnote{We choose to work with $\tilde I_A\equiv I_A-I_\mu$.}
\begin{eqnarray} 
\Delta m^2_{21}\approx4m^2_\nu\left[\left(U^2_{32}-U^2_{31}\right)
\tilde I_\tau + \left(U^2_{12}-U^2_{11}\right)\tilde I_e\right].
\label{eqn:dm12}
\end{eqnarray}
The interesting aspect of this situation is that the (generically) dominant 
first order contribution to $\Delta m^2$ proportional to 
$\tilde I_\tau=\tilde I_\tau^{\rm rg}+\tilde I_\tau^{\rm th}$ vanishes for 
$U^2_{32}=U^2_{31}$, 
i.e. for $s_{13}=\pm(s_{23}/c_{23})(c_{12}\pm s_{12})/(c_{12}\mp s_{12})$, 
a special case of which is the bimaximal mixing solution with $s_{13}=0$ 
(also the second term in eq.~(\ref{eqn:dm12}) then vanishes). This has been 
discussed in refs. \cite{CAESIBNA3,BAROST,CHKRPO} in connection with the 
possibility of realizing the VO solution to the solar neutrino problem, 
within the inversely hierarchical pattern 
$m^2_{\nu_3}\ll m^2_\nu\sim3\times10^{-3}$ eV$^2$. Note that for generic 
mixing angles the RG corrections would always give too large a $\Delta m^2$.

Since the second order correction ($\propto\tilde I^2_\tau$) to $\Delta m^2$ 
is proportional to $m_{\nu_3}$, the right $\Delta m^2$ for the VO solution 
can be obtained by appropriately tuning $m^2_{\nu_3}\simgt0$ and/or due to 
the threshold effects.

As discussed in \cite{CHKRPO}, with $m_{\nu_1}=-m_{\nu_2}$, 
$m^2_{\nu_3}\ll m^2_{\nu_1}\sim3\times10^{-3}$ eV$^2$, the correct
$\Delta m^2$ for the LAMSW (or SAMSW) solutions can be obtained from
the RG running. From eq.~(\ref{eqn:dm12}) we get 
\begin{eqnarray}
\Delta m^2\approx4m^2_\nu\left|s^2_{23}\cos2\theta_{12}+
s_{13}\sin2\theta_{23}\sin2\theta_{12}\right|\tilde I_\tau^{\rm rg}
\end{eqnarray}
and for the LAMSW solution, with $|s_{13}|\sim0.1$, 
$\Delta m^2\sim{\cal O}(10^{-5})$~eV$^2$ can be obtained for 
$\tan\beta\simgt30$ ($\tan\beta\simgt10$ for the SAMSW solution).
For smaller values of $\tan\beta$ one has to investigate potentially
larger threshold corrections, which  give \cite{CHU}
\begin{eqnarray}
&&\Delta m^2\approx4m^2_\nu\left|\sin2\theta_{12}(c_{23}I_{e\mu}^{\rm th}
-s_{23}I_{e\tau}^{\rm th})\right| ~~~~~~~~~{\rm LAMSW}\nonumber\\
&&\Delta m^2\approx4m^2_\nu\left|\cos2\theta_{12}(\tilde I_e^{\rm th} 
-{1\over2}\sin2\theta_{23}
I_{\mu\tau}^{\rm th})\right| ~~~~{\rm SAMSW}.\label{eqn:corrs}
\end{eqnarray}
For $\Delta m^2\sim{\cal O}(10^{-5})$~eV$^2$ one needs therefore corrections
$\tilde I^{\rm th}\sim10^{-3}-10^{-2}$ which is definitely too big a value
for $I_{e\mu}^{\rm th}$ in the MSSM and also impossible for 
$I_{e\tau}^{\rm th}$ and $I_{\mu\tau}^{\rm th}$ (at least within the 
validity of the mass insertion approximation, it would require
$|\delta^{e(\mu)\tau}_{LL}|>1$).

Our discussion here does not cover fully the potential role of quantum 
corrections for the two-fold degeneracy pattern. Several other 
possibilities do exist, depending on the chosen solar solution and we
refer the reader to the literature \cite{CHUPO,CHU,CHWO} for further details.

\section{Conclusions}
\setcounter{equation}{0} 

For every theory of neutrino masses, for a meaningful
comparison with experimental information, it is necessary to discuss quantum
corrections. This is relatively easy if the effective low energy theory is 
the SM or its supersymmetric extension and the neutrino masses enter via the
effective operator (\ref{eqn:dim5op}). In Sections 4 and 5 we reviewed the 
formalism for including quantum correction in those cases. Next we applied 
that formalism to several different classes of the neutrino mass sector at the
scale $M$, hypothetically given by the theory of neutrino masses. They split 
into two broad groups. The first one is characterized by small quantum 
corrections that may eventually be important for precision tests of the future
theory. However, they are irrelevant at the level of present, qualitative
considerations. Here belongs the hierarchical pattern and the partly
degenerate pattern with opposite CP parities of the (almost) degenerate
neutrinos. For those mass matrices, quantum corrections cannot substantially
alter the structures present at the scale $M$, so the agreement with 
experimental data  has to be assured by the boundary values at $M$. Equal
masses $m_{\nu_1}=-m_{\nu_2}$, at the scale $M$ are possible. Quantum 
corrections can explain the observed $\Delta m^2$.

The other group consists of partly degenerate pattern with the same CP 
parities of the (almost) degenerate neutrinos and of the (approximately) 
degenerate structures. Large quantum corrections can originate either from 
the RG evolution or from low energy threshold corrections (in the MSSM). 
They never change qualitatively the mass eigenvalue pattern, although they 
may explain their observed splitting. However, large quantum corrections 
always lead to a ``fixed point'' relation for the mixing angles. The 
sufficient condition is that for at least one pair of neutrinos 
$(m_{\nu_a}+m_{\nu_b})/(m_{\nu_a}-m_{\nu_b})\gg1$. It is interesting that 
the transition from small (qualitatively irrelevant) quantum corrections to 
the ``fixed point'' behaviour is very abrupt. So, to a good approximation 
there are those two and only two physical situations.

A fixed point relation is always one equation for three angles (if CP is
conserved) and makes their low energy values dependent on only two boundary 
conditions for the angles at the scale $M$. If the dominant quantum 
corrections come from the RG evolution (i.e. originate from the large $\tau$
Yukawa coupling) or also from several configurations of the low energy 
threshold corrections, the ``fixed point'' relation links small 
$\theta_{13}$ angle (constrained by the CHOOZ experiment) to a small angle
responsible for the mixing of solar neutrinos. If presently favoured
bimaximal mixing was confirmed by future experimental data, all mass patterns
leading to large RG corrections and the regions of the MSSM parameter space
leading to the same ``fixed point'' relation would be ruled out. 

There are,
however, other ``fixed point'' relations, generated by some special sfermion
mass configurations and/or by flavour non-conserving effects in the slepton
sector, that are consistent with bimaximal mixing and small $\theta_{13}$ 
angle. Such solutions are phenomenologically interesting as, at the same time, 
they explain the observed mass squared differences, as the effect of quantum 
corrections, with degenerate spectrum at the scale $M$.\footnote{It has 
also been discussed in the literature \cite{CAESIBNA1,CAESIBNA2} that 
quantum corrections generated in the full theory above the scale $M$,
can break the degeneracy already at the scale $M$ and lead to correct mass
squared differences and mixing angles at the electroweak scale.
For an interesting link between neutrino masses and flavour changing 
processes see ref. \cite{LAMASA}.}
However, the simple Ansatz at the scale $M$ needs a deeper theoretical
justification.

Quantum corrections do not explain the origin of the neutrino masses and do
not replace its theory. Nevertheless, they are important piece of the overall
picture. They will constrain strongly the acceptable mass structures once the 
experimental ambiguities are resolved.

\vskip1.0cm

\vskip1.0cm
\noindent {\bf Acknowledgments}
\vskip 0.3cm
\noindent 
The work of P.H.Ch. was supported partially by the EC Contract 
HPRN-CT-2000-00148 for years 2000-2004 and by the Polish State Committee 
for Scientific Research grant 2 P03B 060 18 for years 2000-2001.
The work of S.P. was supported partially by the EC Contract 
HPRN-CT-2000-00152 for years 2000-2004 and by the Polish State Committee 
for Scientific Research grant  5 P03B 119 20  for years 2001-2002. 
P.H.Ch. Would also like to thank the CERN Theory group
for hospitality during the writing of this article.

\renewcommand{\thesection}{Appendix~\Alph{section}}
\renewcommand{\theequation}{\Alph{section}.\arabic{equation}}

\setcounter{section}{0}
\setcounter{equation}{0}

\section{}

In Section 3, instead of the familiar Dirac bispinors we used the 
two-component Weyl spinor notation. It is particularily convenient for 
dealing with Majorana particles. Here we explain this notation shortly. More 
details
can be found in modern textbooks on QFT (see e.g. \cite{POKBOOK,PESIO}).

In four dimensions, the Lorentz group or more precisely its covering group
$SL(2,C)$ has two non-equivalent complex two-dimensional representations
denoted as $(1/2,0)$ and $(0,1/2)$. The Grassmann fields 
(or fermionic field operators) transforming according to
these representations are conventionally written as $\lambda_\alpha$ and
$\bar\chi^{\dot\alpha}$ and called\footnote{Somewhat incorrectly; 
properly they should be called left- and right-chiral.} left- and 
right-handed spinors, respectively. Since the complex conjugation of
a left-handed spinor $\lambda_\alpha$ transforms as a right-handed one
($(\lambda_\alpha)^\ast\sim\bar\lambda_{\dot\alpha}$),
the  fermion content of any Lagrangian can be specified by listing only
the left-handed spinors used for its construction.

If the two left-handed fields $\lambda_\alpha$ and $\chi_\beta$ transform 
as representations $R$ and $R^\ast$, respectively under the final unbroken 
symmetry group (global or local) of the theory, they
can be combined to form a Dirac bispinor:
\begin{eqnarray}
\psi_{(\lambda)}
=\left(\matrix{\lambda_\alpha\cr\bar\chi^{\dot\beta}}\right), ~~~~~~
\bar\psi_{(\lambda)}=\left(\chi^\alpha,\bar\lambda^{\dot\beta}\right),
\label{eqn:A1}
\end{eqnarray}
transforming as $R$ and $R^\ast$, respectively.
The raising and lowering of Weyl spinor indices is done with the help
of the antisymmetric tensors $\epsilon^{\alpha\beta}$, 
$\epsilon_{\alpha\beta}$,
$\epsilon_{\dot\alpha\dot\beta}$ and $\epsilon^{\dot\alpha\dot\beta}$:
\begin{eqnarray}
\lambda^\alpha=\epsilon^{\alpha\beta}\lambda_\beta, ~~
\lambda_\alpha=\lambda^\beta\epsilon_{\beta\alpha}, ~~
\bar\chi_{\dot\alpha}=\epsilon_{\dot\alpha\dot\beta}\bar\chi_{\dot\beta}, ~~
\bar\chi^{\dot\alpha}=\bar\chi_{\dot\beta}\epsilon^{\dot\beta\dot\alpha}.
\end{eqnarray}
The two kinetic terms for $\lambda$ and $\chi$ can be
then rewritten in the familiar form
\begin{eqnarray}
{\cal L}_{\rm kin}=i\bar\lambda\bar\sigma^\mu\partial_\mu\lambda +
i\bar\chi\bar\sigma^\mu\partial_\mu\chi =
i\bar\lambda\bar\sigma^\mu\partial_\mu\lambda +
i\chi\sigma^\mu\partial_\mu\bar\chi + {\rm (total ~der)}
=i\bar\psi_{(\lambda)}\gamma^\mu\partial_\mu\psi_{(\lambda)}\label{eqn:kin}
\end{eqnarray}
where the Dirac matrices $\gamma^\mu$ in the Weyl representation are
constructed as
\begin{eqnarray}
\gamma^\mu=\left(\matrix{0&\sigma^\mu_{\alpha\dot\beta}\cr
\bar\sigma^{\mu\dot\alpha\beta}&0}\right)
\end{eqnarray}
with $\sigma^\mu\equiv\left(I,\mbox{\boldmath$\sigma$}\right)$, 
$\bar\sigma^\mu\equiv\left(I,-\mbox{\boldmath$\sigma$}\right)$ 
($\mbox{\boldmath$\sigma$}$'s are the Pauli matrices).
For such a pair of Weyl fields also a Dirac mass term can be constructed
\begin{eqnarray}
{\cal L}_{\rm mass}= -m\left(\lambda^\alpha\chi_\alpha
+ m\bar\lambda_{\dot\alpha}\bar\chi^{\dot\alpha}\right) 
=-m\bar\psi_{(\lambda)}\psi_{(\lambda)}\label{eqn:massterm}
\end{eqnarray}
If the field $\lambda$ ($\chi$) has no left-handed partner transforming in the
complex conjugate representation $R^\ast$ ($R$), it is convenient to introduce
chiral Dirac bispinors
\begin{eqnarray}
\psi_{(\lambda)L}=\left(\matrix{\lambda_\alpha\cr0}\right), ~~~
\psi_{(\chi)R}=\left(\matrix{0\cr\bar\chi^{\dot\alpha}}\right).
\end{eqnarray}
For chiral Dirac bispinors e.g. $\psi_{(\lambda)L}=P_L\psi_{(\lambda)L}$ etc.,
where $P_L\equiv(1-\gamma^5)/2$. Note that (see eq.(\ref{eqn:A1})),
\begin{eqnarray}
\overline{\psi_{(\lambda)L}} = \left(\matrix{0&\bar\lambda_{\dot\alpha}}\right),
\phantom{aaaa}
\overline{\psi_{(\chi)R}} = \left(\matrix{\chi^\alpha&0}\right).
\end{eqnarray}

The typical Yukawa coupling of a scalar field $\phi$ in the 
representation $R_\phi$ and two left-handed Weyl spinors $\lambda$ and $\chi$ 
transforming as representations $R_\lambda$ and $R_\chi$, respectively 
(such that $1\subset R_\phi\times R_\lambda\times R_\chi$) can be written 
as (we omit the Clebsch-Gordan coefficients)
\begin{eqnarray}
&&{\cal L}_{\rm Yuk}=-Y\phi\lambda\chi
-Y^\ast\phi^\dagger\bar\lambda\bar\chi\nonumber\\
&&\phantom{aaaa}=
-Y\phi\overline{\psi_{(\chi)R}}\psi_{(\lambda)L}
-Y^\ast\phi^\dagger\overline{\psi_{(\lambda)L}}\psi_{(\chi)R}
\label{eqn:lryuk}
\end{eqnarray}
The Yukawa part of the SM Lagrangian (\ref{eqn:smlyuk}) and the SM mass
terms (\ref{eqn:smlmass}) are the  example
of (\ref{eqn:lryuk}) and (\ref{eqn:massterm}), respectively,
with fields $u^c$, $d^c$ and $e^c$ playing the role of
$\chi$, and $q$ and $l$ (or $u$, $d$ and $e$) playing the role of $\lambda$.

Finally, Weyl spinor fields $\lambda_\alpha$ which are singlets of all
unbroken symmetries of the theory can form 4-component Majorana bispinors
\begin{eqnarray}
\psi_{(\lambda){\rm Maj}}=
\left(\matrix{\lambda^\alpha\cr\bar\lambda^{\dot\beta}}\right).
\end{eqnarray}
Of course in this case 
$\psi_{(\lambda)}=C\bar\psi_{(\lambda)}^T\equiv\psi^c_{(\lambda)}$ which
means that the field is self-conjugate. One also has (up to raising or
lowering indices)
\begin{eqnarray}
\lambda=\psi_{(\lambda)L}=\overline{\psi_{(\lambda)R}}
{\rm ~~~~and} ~~~~
\bar\lambda=\psi_{(\lambda)R}=\overline{\psi_{(\lambda)L}}.
\end{eqnarray}
For such a field a Majorana mass term can be formed
\begin{eqnarray}
&&{\cal L}_{\rm Maj}=-{1\over2}m(\lambda\lambda+\bar\lambda\bar\lambda)
=-{1\over2}m\left(\overline{\psi_{(\lambda)R}}\psi_{(\lambda)L}
+\overline{\psi_{(\lambda)L}}\psi_{(\lambda)R}\right)\nonumber\\
&&\phantom{aaaa}\equiv-{1\over2}m\bar\psi_{(\lambda)}\psi_{(\lambda)}
\equiv-{1\over2}m\psi_{(\lambda)}^TC\psi_{(\lambda)}
\end{eqnarray}
where $C$  is the charge conjugation matrix. 
The Majorana mass term (\ref{eqn:Majmass}) is precisely of this form
with $\nu^c$ playing the role of $\lambda$ and the Yukawa coupling
(\ref{eqn:Yukneu}) is usually written as 
\begin{eqnarray}
\Delta{\cal L}_{\rm Yuk}=-\epsilon_{ij}H_i
\overline{\psi_{(\nu^{c})R}}\mbox{\boldmath$Y$}_\nu
\psi_{(l)L} -\epsilon_{ij}H_i^\ast
\overline{\psi_{(\lambda)L}}\mbox{\boldmath$Y$}_\nu^\dagger
\psi_{(\nu^{c})R}
\end{eqnarray}

\setcounter{equation}{0}
\section{}

In this Appendix we recall the well known RGEs for Yukawa coupling matrices 
defined in eq.~(\ref{eqn:smlyuk}) in the SM and in the MSSM.
In the SM they read \cite{BABU,BABEOH1}:
\begin{eqnarray}
&&{d\over dt}\mbox{\boldmath$Y$}_u = 
\mbox{\boldmath$Y$}_u\left[-8g^2_3-{9\over4}g^2_2-{17\over12}g^2_Y + T
+{3\over2}\left(\mbox{\boldmath$Y$}_u^\dagger\mbox{\boldmath$Y$}_u
-\mbox{\boldmath$Y$}_d^\dagger\mbox{\boldmath$Y$}_d\right)\right],
\phantom{aaa}\nonumber\\
&&{d\over dt}\mbox{\boldmath$Y$}_d = 
\mbox{\boldmath$Y$}_d\left[-8g^2_3-{9\over4}g^2_2-{5\over12}g^2_Y + T
+{3\over2}\left(\mbox{\boldmath$Y$}_d^\dagger\mbox{\boldmath$Y$}_d
-\mbox{\boldmath$Y$}_u^\dagger\mbox{\boldmath$Y$}_u\right)\right],
\phantom{aaa}\label{eqn:dpoptY_sm}\\
&&{d\over dt}\mbox{\boldmath$Y$}_e = 
\mbox{\boldmath$Y$}_e\left[-{9\over4}g^2_2-{15\over4}g^2_Y + T
+{3\over2}\mbox{\boldmath$Y$}_e^\dagger\mbox{\boldmath$Y$}_e\right],
\phantom{aaa}\nonumber
\end{eqnarray}
where 
\begin{eqnarray}
t\equiv{1\over16\pi^2}\ln\left({Q\over M_Z}\right),
\end{eqnarray}
\begin{eqnarray}
T\equiv{\rm Tr}\left(3\mbox{\boldmath$Y$}_u^\dagger\mbox{\boldmath$Y$}_u
+3\mbox{\boldmath$Y$}_d^\dagger\mbox{\boldmath$Y$}_d
+\mbox{\boldmath$Y$}_e^\dagger\mbox{\boldmath$Y$}_e\right)
\label{eqn:Tsm}
\end{eqnarray}
and the gauge couplings $g_3$, $g_2$ and $g_Y$ evolve according to
\begin{eqnarray}
{d\over dt}g_i=b_ig_i^3\phantom{aaa}i=3,2,Y
\end{eqnarray}
with $b_3=-7$, $b_2=-19/6$ and $b_Y=41/6$.

In the MSSM one finds instead:
\begin{eqnarray}
{d\over dt}\mbox{\boldmath$Y$}_u = 
\mbox{\boldmath$Y$}_u\left[-{16\over3}g^2_3-3g^2_2-{13\over9}g^2_Y 
+{\rm Tr} \left(3\mbox{\boldmath$Y$}_u^\dagger\mbox{\boldmath$Y$}_u\right)
+3\mbox{\boldmath$Y$}_u^\dagger\mbox{\boldmath$Y$}_u
+\mbox{\boldmath$Y$}_d^\dagger\mbox{\boldmath$Y$}_d\right],
\phantom{aaaaaa}\nonumber\\
{d\over dt}\mbox{\boldmath$Y$}_d = 
\mbox{\boldmath$Y$}_d\left[-{16\over3}g^2_3-3g^2_2-{7\over9}g^2_Y 
+{\rm Tr}\left(3\mbox{\boldmath$Y$}_d^\dagger\mbox{\boldmath$Y$}_d +
\mbox{\boldmath$Y$}_e^\dagger\mbox{\boldmath$Y$}_e\right)
+3\mbox{\boldmath$Y$}_d^\dagger\mbox{\boldmath$Y$}_d
+\mbox{\boldmath$Y$}_u^\dagger\mbox{\boldmath$Y$}_u\right],
\label{eqn:dpoptY_mssm}\\
{d\over dt}\mbox{\boldmath$Y$}_e = 
\mbox{\boldmath$Y$}_e\left[-3g^2_2-3g^2_Y + {\rm Tr}\left(
3\mbox{\boldmath$Y$}_d^\dagger\mbox{\boldmath$Y$}_d +
\mbox{\boldmath$Y$}_e^\dagger\mbox{\boldmath$Y$}_e\right)
+3\mbox{\boldmath$Y$}_e^\dagger\mbox{\boldmath$Y$}_e\right],
\phantom{aaaaaaaaaaaaaaa}\nonumber
\end{eqnarray}
and the factors $b_i$ change to $b_3=-3$, $b_2=+1$ and $b_Y=11$.

For completeness we give here also the RGEs above the scale $M$
i.e. for the theory whose set of fermion fields includes additional
three $SU_L(2)\times U_Y(1)$ singlet neutrino fields $\nu^{cA}$ 
($A=1,2,3$) and whose Lagrangian is identical to the one for the
effective theory valid below $M$ except for the Yukawa interaction
(\ref{eqn:Yukneu}) and the Majorana mass terms (\ref{eqn:Majmass}).

If the theory above the $M$ scale extends the
SM, then  $T$ given in eq. (\ref{eqn:Tsm}) has to be replaced by
\cite{GRLI,HAOKSU,HAMAOKSU1}
\begin{eqnarray}
T\equiv{\rm Tr}\left(
3\mbox{\boldmath$Y$}_u^\dagger\mbox{\boldmath$Y$}_u
+3\mbox{\boldmath$Y$}_d^\dagger\mbox{\boldmath$Y$}_d
+\mbox{\boldmath$Y$}_e^\dagger\mbox{\boldmath$Y$}_e
+\mbox{\boldmath$Y$}_\nu^\dagger\mbox{\boldmath$Y$}_\nu\right),
\end{eqnarray}
the last equation in (\ref{eqn:dpoptY_sm}) should be replaced by
\begin{eqnarray}
{d\over dt}\mbox{\boldmath$Y$}_e=\mbox{\boldmath$Y$}_e
\left[-{9\over4}g^2_2-{15\over4}g^2_Y+T+{3\over2}
\left(\mbox{\boldmath$Y$}_e^\dagger\mbox{\boldmath$Y$}_e
-\mbox{\boldmath$Y$}_\nu^\dagger\mbox{\boldmath$Y$}_\nu\right)\right]
\end{eqnarray}
and the neutrino Yukawa matrix RGE reads:
\begin{eqnarray}
{d\over dt}\mbox{\boldmath$Y$}_\nu=\mbox{\boldmath$Y$}_\nu
\left[-{9\over4}g^2_2-{3\over4}g^2_Y+T-{3\over2}
\left(\mbox{\boldmath$Y$}_e^\dagger\mbox{\boldmath$Y$}_e
-\mbox{\boldmath$Y$}_\nu^\dagger\mbox{\boldmath$Y$}_\nu\right)\right].
\end{eqnarray}
In addition, the Majorana mass matrix also runs \cite{CAESIBNA1}:
\begin{eqnarray}
{d\over dt}\mbox{\boldmath$M$}_{\rm Maj}^{KL}
=\mbox{\boldmath$M$}_{\rm Maj}^{KJ}
\left(\mbox{\boldmath$Y$}_\nu\mbox{\boldmath$Y$}_\nu^\dagger\right)^{LJ}
+\left(\mbox{\boldmath$Y$}_\nu\mbox{\boldmath$Y$}_\nu^\dagger\right)^{KJ}
\mbox{\boldmath$M$}_{\rm Maj}^{JL}.
\end{eqnarray}

If the low energy theory is the MSSM, then \cite{HAOKSU,HAMAOKSU1}
in the first equation of (\ref{eqn:dpoptY_mssm}) 
\begin{eqnarray}
{\rm Tr}\left(3\mbox{\boldmath$Y$}_u^\dagger\mbox{\boldmath$Y$}_u\right)
\rightarrow
{\rm Tr}\left(3\mbox{\boldmath$Y$}_u^\dagger\mbox{\boldmath$Y$}_u
+\mbox{\boldmath$Y$}_\nu^\dagger\mbox{\boldmath$Y$}_\nu\right),
\end{eqnarray}
the last equation of (\ref{eqn:dpoptY_mssm}) is replaced by
\begin{eqnarray}
{d\over dt}\mbox{\boldmath$Y$}_e=\mbox{\boldmath$Y$}_e
\left[-3g^2_2-3g^2_Y+{\rm Tr}\left(
3\mbox{\boldmath$Y$}_d^\dagger\mbox{\boldmath$Y$}_d
+\mbox{\boldmath$Y$}_e^\dagger\mbox{\boldmath$Y$}_e\right)
+3\mbox{\boldmath$Y$}_e^\dagger\mbox{\boldmath$Y$}_e
+\mbox{\boldmath$Y$}_\nu^\dagger\mbox{\boldmath$Y$}_\nu\right]
\end{eqnarray}
and the neutrino Yukawa matrix RGE reads:
\begin{eqnarray}
{d\over dt}\mbox{\boldmath$Y$}_\nu=\mbox{\boldmath$Y$}_\nu
\left[-3g^2_2-g^2_Y+{\rm Tr}\left(
3\mbox{\boldmath$Y$}_u^\dagger\mbox{\boldmath$Y$}_u
+\mbox{\boldmath$Y$}_\nu^\dagger\mbox{\boldmath$Y$}_\nu\right)
+3\mbox{\boldmath$Y$}_e^\dagger\mbox{\boldmath$Y$}_e
+\mbox{\boldmath$Y$}_\nu^\dagger\mbox{\boldmath$Y$}_\nu\right].
\end{eqnarray}
Finally, the Majorana mass matrix running is dictated by \cite{CAESIBNA2}:
\begin{eqnarray}
{d\over dt}\mbox{\boldmath$M$}_{\rm Maj}^{KL}
=2\mbox{\boldmath$M$}_{\rm Maj}^{KJ}
\left(\mbox{\boldmath$Y$}_\nu\mbox{\boldmath$Y$}_\nu^\dagger\right)^{LJ}
+2\left(\mbox{\boldmath$Y$}_\nu\mbox{\boldmath$Y$}_\nu^\dagger\right)^{KJ}
\mbox{\boldmath$M$}_{\rm Maj}^{JL}.
\end{eqnarray}


\vskip1.0cm

\begin{thebibliography}{99}

\bibitem{NI} K. Nishikawa, plenary talk at the {\sl International EPS 
             Conference on High Energy Physics}, Budapest, Hungary,
             July 2001.

\bibitem{ALFE} G. Altarelli and F. Feruglio,  {\sl Phys. Lett.} {\bf B439}
               (1998), 112, {\sl JHEP} {\bf 9811:021} (1998), 
               {\sl Phys. Rept.} {\bf 320} (1999), 295;
               G. Altarelli, F. Feruglio and I. Masina, {\sl Phys. Lett.}
               {\bf B472} (2000), 382.

\bibitem{PDG} The Particle Data Group, {\sl Eur. Phys. J.} {\bf C15} (2000), 1.

\bibitem{CA} C. Giunti, C.W. Kim, J.A. Lee and U.W. Lee, {\sl Phys. Rev.} 
             {\bf D48} (1993), 4310; J. Rich, {\sl Phys. Rev.} 
             {\bf D48} (1993), 4318; C. Cardall {\sl Phys. Rev.} 
             {\bf D61} (2000), 073006; M. Zra\l ek, {\sl Acta Phys. Pol.}
             {\bf B29} (1998), 3925.

\bibitem{BIGIGR} S.M. Bilenkii, C. Giunti and W. Grimus, {\sl Prog. Part.
                 Nucl. Phys.} {\bf 43} (1999), 1.

\bibitem{MISMWO} S.P. Mikheyev and A.Yu. Smirnov, {\sl Sov. J. Nucl. Phys.}
                 {\bf 42} (1985), 913; L.F. Wolfenstein, {\sl Phys. Rev.} 
                 {\bf D17} (1978), 2369.

\bibitem{SUPERK} Y. Fukuda et al.  (The Superkamiokande Collaboration),
                 {\sl Phys. Rev. Lett.} {\bf 81} (1998), 1562.

\bibitem{SUPERK_NOs} S. Fukuda  et al. (The Superkamiokande Collaboration), 
                     {\sl Phys. Rev. Lett.} {\bf 85} (2000), 3999.

\bibitem{BEGRVO} C. Bemporad, G. Gratta and P. Vogel, hep-ph/0107277.   

\bibitem{LSND} C. Athanassopoulos et al.  (The LSND Collaboration),
               {\sl Phys. Rev. Lett.} {\bf 77} (1996), 3082, 
               {\bf 81} (1998), 1774. 

\bibitem{SNO} Q.R. Ahmad et al. (The SNO Collaboration), 
              {\sl Phys. Rev. Lett.} {\bf 87} (2001), 071301.

\bibitem{SUPERK_sol} S. Fukuda et al.  (The Superkamiokande Collaboration),
                     {\sl Phys. Rev. Lett.} {\bf 86} (2001), 5651, 5656.

\bibitem{CHOOZ} M. Apollonio et al. (The CHOOZ Collaboration),
                {\sl Phys. Lett.} {\bf B466} (1999), 415.

\bibitem{MANASA} Z. Maki, M. Nakagawa and S. Sakata, {\sl Prog. Theor. Phys.}
                 {\bf 28} (1962), 870.

\bibitem{BETA0NIU} L. Baudis et al. (Heidelberg-Moscow Collaboration),
                   {\sl Phys. Rev. Lett.} {\bf 83} (1999), 41;
                   A. Dietz (Heidelberg-Moscow Collaboration), talk presented
                   at the $3^{rd}$ {\sl International Conference on Dark 
                   Matter in Astro and Particle Physics},
                   Heidelberg, Germany, Jul 2000 (hep-ph/0103062).

\bibitem{POKBOOK} S. Pokorski, {\sl Gauge Field Theories}, 2nd edition,
                  Cambridge University Press, Cambridge, 2000, Chapters 
                  1 and 12. 

\bibitem{GERASL} M. Gell-Mann, P. Ramond and R. Slansky, in proceedings of the
                 {\sl Supergravity Stony Brook Workshop}, New York, 1979,
                 eds. P. Van Nieuwenhuizen and D. Freedman (North-Holland,
                 Amsterdam); T. yanagida, in proceedings of the {\sl Workshop
                 on Unified Theories and Baryon Number in the Universe}, 
                 Tsukuba, 1979, eds. A. Sawada and A. Sugamoto, KEK Report
                 No. 79-18.

\bibitem{MOH} R.N. Mohapatra, preprint UMDPP-02-002, 2001 (hep-ph/0107264).

\bibitem{WEI} S. Weinberg, {\sl Phys. Rev. Lett.} {\bf 43} (1979), 1566.

\bibitem{BALE} K.S. Babu and C.N. Leung,  preprint OSU-HEP-01-02, UDHEP-02-01,
               hep-ph/0106054.

\bibitem{HE} R. Hempfling, {\sl Nucl. Phys.} {\bf B478} (1996), 3; 
             E. Nardi, {\sl Phys. Rev.} {\bf D55} (1997), 5772;
             E.J. Chun, S.K. Kang, C.W. Kim and U.W. Lee, {\sl Nucl. Phys.} 
             {\bf B544} (1999), 89; E.J. Chun and S.K. Kang {\sl Phys. Rev.}
             {\bf D61} (2000), 075012.

\bibitem{CHEILI} T.P. Cheng, E. Eichten and L.-F. Li, {\sl Phys. Rev.}
                 {\bf D9} (1974) 2259.

\bibitem{MAVA} M.E. Macha\v cek and M.T. Vaughn {\sl Nucl. Phys.} {\bf B236}
               (1984), 221.

\bibitem{BABEOH1} V. Barger, M.S. Berger and P. Ohmann, {\sl Phys. Rev.} 
                 {\bf D47} (1993), 1093.

\bibitem{BABU} K.S. Babu, {\sl Z. Phys.} {\bf C35} (1987), 69.

\bibitem{OLPO} M. Olechowski and S. Pokorski, {\sl Phys. Lett.} {\bf 231B}
               (1989) 165; {\sl ibid.} {\bf 257B} (1991), 388.

\bibitem{BABEOH2} V. Barger, M.S. Berger and P. Ohmann, {\sl Phys. Rev.} 
                 {\bf D47} (1993), 2038.

\bibitem{WO} L.F. Wolfenstein {\sl Phys. Rev. Lett.} {\bf 51} (1983), 1945.

\bibitem{ANDRKELIRA} S. Antusch, M. Drees, J. Kersten, M. Lindner and
                     M. Ratz, preprint TUM-HEP-424/01 (hep-ph/0108005).

\bibitem{WE} C. Wetterich, {\sl Nucl. Phys.} {\bf B187} (1981), 343;
             A. Ioannisian {\sl Sov. J. Nucl. Phys.} {\bf 51} (1990), 511.

\bibitem{CHPL} P.H. Chankowski and Z. P\l uciennik, {\sl Phys. Lett.} 
               {\bf B316} (1993), 312.

\bibitem{BALEPA} K.S. Babu, C.N. Leung and J. Pantaleone,  {\sl Phys. Lett.} 
                 {\bf B319} (1993), 191. 

\bibitem{CH} P.H. Chankowski, {\sl Phys. Rev.} {\bf D41} (1990), 2877.

\bibitem{ELLO} J. Ellis and S. Lola, {\sl Phys. Lett.} {\bf B458} (1999), 310.

\bibitem{HAMAOKSU1} N. Haba, Y. Matsui, N. Okamura and M. Sugiura,
                   {\sl Eur. Phys. J.} {\bf C10} (1999), 677.

\bibitem{CHKRPO} P.H. Chankowski, W. Kr\'olikowski and S. Pokorski,
                 {\sl Phys. Lett.} {\bf B473} (2000), 109.

\bibitem{CAESIBNA4} J.A. Casas, J.R. Espinosa, A. Ibarra and I. Navarro,
                    {\sl Nucl. Phys.} {\bf B573} (2000), 652.

\bibitem{TA} M. Tanimoto, {\sl Phys. Lett.} {\bf B360} (1995), 41

\bibitem{HAOKSU} N. Haba, N, Okamura and M. Sugiura, {\sl Prog. Theor. Phys.}
                 {\bf 103} (2000), 367.

\bibitem{BADIMOPA1} K.R.S. Balaji, A. Dighe, R.N. Mohapatra and M.K. Parida, 
                   {\sl Phys. Rev. Lett.} {\bf 84} (2000), 5034.

\bibitem{BAMOPAPA} K.R.S. Balaji, R.N. Mohapatra, M.K. Parida and E.A. Paschos,
                   {\sl Phys. Rev.} {\bf D63}:113002 (2001).

\bibitem{ELLELONA} J. Ellis, G.K. Leontaris, S. Lola and D.V. Nanopoulos,
                   {\sl Eur. Phys. J.} {\bf C9} (1999) 389.

\bibitem{LO} S. Lola, proceedings of the 6$^{th}$ Hellenic School and 
             Workshop on Elementary Particle Physics, Corfu, Greece, 
             September 1998 (hep-ph/9903203).

\bibitem{CAELLOWA} M. Carena, J. Ellis, S. Lola, C.E.M. Wagner,
                   {\sl Eur. Phys. J.} {\bf C12} (2000), 507.

\bibitem{MA} E. Ma, {\sl J. Phys.} {\bf G25} (1999), L97-L100.

\bibitem{HAMAOK} N. Haba, Y, Matsui and N. Okamura,  {\sl Prog. Theor. Phys.}
                 {\bf 103} (2000), 807; T.K. Kuo, J. 
                 Pantaleone and G.-H. Wu.

\bibitem{KUPAWU} T.K. Kuo, J. Pantaleone and G.-H. Wu, hep-ph/0103131.

\bibitem{HAMAOKSU2} N. Haba, Y, Matsui, N, Okamura and M. Sugiura,  
                   {\sl Prog. Theor. Phys.} {\bf 103} (2000), 145. 

\bibitem{BAROST} R. Barbieri, G.G. Ross and A. Strumia,  
                 {\sl JHEP} {\bf 9910:020} (1999).

\bibitem{CAESIBNA3} J.A. Casas, J.R. Espinosa, A. Ibarra and I. Navarro,
                    {\sl JHEP} {\bf 9909}:015 (1999).

\bibitem{DIJO} A.S. Dighe and A.S. Joshipura, preprint CERN-TH-2000-301
               (hep-ph/0010079).

\bibitem{CAESIBNA1} J.A. Casas, J.R. Espinosa, A. Ibarra and I. Navarro,
                    {\sl Nucl. Phys.} {\bf B556} (1999), 3.

\bibitem{CAESIBNA2} J.A. Casas, J.R. Espinosa, A. Ibarra and I. Navarro,
                    {\sl Nucl. Phys.} {\bf B569} (2000), 82.

\bibitem{HAOK} N. Haba and N. Okamura, {\sl Eur. Phys. J.} {\bf C14} (2000),
               347.

\bibitem{BADIMOPA2} K.R.S. Balaji, A. Dighe, R.N. Mohapatra and M.K. Parida, 
                    {\sl Phys. Lett.} {\bf B481} (2000), 33.

\bibitem{CHWO} P.H. Chankowski and P. Wasowicz, preprint IFT-01/28,
               (hep-ph/0110237).

\bibitem{SCHIFF} L.I. Schiff, {\sl Quantum Mechanics}, 3rd edition, 
                 McGraw-Hill Book Company, New York, 1968.

\bibitem{CHIOPOVA} P.H. Chankowski, A. Ioannisian, S. Pokorski and J.W.F.
                   Valle, {\sl Phys. Rev. Lett.} {\bf 86} (2001), 3488.

\bibitem{DUPOSA} E. Dudas, S. Pokorski and C.A. Savoy, {\sl Phys. Lett.}
                 {\bf B369} (1996), 255; E. Dudas, Ch. Grojean, S. Pokorski 
                 and C.A. Savoy {\sl Nucl. Phys.} {\bf B481} (1996), 85;
                 A.G. Cohen, D.B. Kaplan and A.E. Nelson, {\sl Phys. Lett.} 
                 {\bf B388} (1996), 588; Ph. Brax and C.A. Savoy, 
                 {\sl J. High. Energy Phys.} {\bf 0007}:048 (2000).

\bibitem{CHU} E. J. Chun, {\sl Phys. Lett.} {\bf B505} (2001) 155.

\bibitem{CHUPO} E.J. Chun and S. Pokorski, {\sl Phys. Rev.} {\bf D62}:053001
                (2000).

\bibitem{GAGAMASI} F. Gabbiani, E. Gabrieli, A. Masiero and L. Silvestrini,
                   {\sl Nucl. Phys.} {\bf B447} (1996), 321; see also, M. 
                   Misiak, S. Pokorski and J. Rosiek, in {\sl Heavy Flavours 
                   II}, eds. A.J. Buras and M. Lindner, World Scientific
                   Publishing Co., Singapore 1998.

\bibitem{LAMASA} S. Lavignac, I. Masina and C.A. Savoy, preprint 
                 SACLAY-T01/066 (hep-ph/0106245).

\bibitem{PESIO} M.E. Peskin and D.V. Schroeder, {\sl An Introduction to 
                Quantum Field Theory}, Addison Wesley, Reading Massachussets
                1997.
                
\bibitem{GRLI} B. Grzadkowski and M. Lindner, {\sl Phys. Lett.} {\bf B193}
               (1987), 71; Y.F. Pirogov and O. Zenin, hep-ph/9808396.
                
\end{thebibliography}
\end{document}